\documentclass[prl,twocolumn,superscriptaddress]{revtex4-2}

\usepackage[utf8]{inputenc}
\usepackage[shortcuts]{extdash}
\usepackage{hyperref}
\usepackage{physics}
\usepackage{amssymb,amsmath,dsfont}
\usepackage{graphicx}
\usepackage{natbib}
\usepackage{color}
\usepackage{bm}
\usepackage{csvsimple}
\usepackage{siunitx,booktabs}
\usepackage[caption=false]{subfig}
\usepackage{tabularx}
\usepackage{ulem}
\newcommand{\balpha}{\ensuremath{{\bm \alpha}}}
\newcommand{\btalpha}{\ensuremath{\tilde{\bm \alpha}}}
\newcommand{\bmu}{\ensuremath{{\bm \mu}}}
\newcommand{\mean}[1]{\langle {#1} \rangle}
 
\newcommand{\entVN}{H}
\newcommand{\id}{\mathds{1}}
\newcommand*{\chat}[1]{\hspace{0.15em}\hat{\phantom{#1}}\kern-0.8em #1} 
\renewcommand{\t}[1]{\mathrm{#1}}

\newcommand{\ii}{\mathrm{i}}
\definecolor{mygreen}{rgb}{0.03, 0.47, 0.19}

\begin{document}

\title{
Automated  design of quantum optical experiments\\for device-independent quantum key distribution
}

\author{Xavier Valcarce}
\email{xavier.valcarce@ipht.fr}
\affiliation{Universit\'e Paris-Saclay, CEA, CNRS, Institut de physique th\'eorique, 91191, Gif-sur-Yvette, France}
\author{Pavel Sekatski}
\affiliation{Department of Applied Physics, University of Geneva, 1205 Geneva, Switzerland}
\author{Elie Gouzien}
\affiliation{Universit\'e Paris-Saclay, CEA, CNRS, Institut de physique th\'eorique, 91191, Gif-sur-Yvette, France}
\author{Alexey Melnikov}
\affiliation{Terra Quantum AG, 9000 St Gallen, Switzerland}
\author{Nicolas Sangouard}
\affiliation{Universit\'e Paris-Saclay, CEA, CNRS, Institut de physique th\'eorique, 91191, Gif-sur-Yvette, France}
\begin{abstract}
    Device-independent quantum key distribution (DIQKD) reduces the vulnerability to side-channel attacks of standard QKD protocols by removing the need for characterized quantum devices. The higher security guarantees come however, at the price of a challenging implementation. Here, we tackle the question of the conception of an experiment for implementing DIQKD with photonic devices. We introduce a technique combining reinforcement learning, optimisation algorithm and a custom efficient simulation of quantum optics experiments to automate the design of photonic setups maximizing a given function of the measurement statistics. Applying the algorithm to DIQKD, we get unexpected experimental configurations leading to high key rates and to a high resistance to loss and noise. These configurations might be helpful to facilitate a first implementation of DIQKD with photonic devices and for future developments targeting improved performances.
\end{abstract}
\maketitle
\paragraph{Introduction --} In quantum key distribution (QKD)~\cite{Bennett84,Ekert91}, two separated parties connected by a public quantum channel, Alice \& Bob, aim at expanding a string of private random bits, i.e.\@ a key. The secrecy and correctness of this key, i.e.\@ the guarantee that an eavesdropper, Eve, who may control the quantum channel has no information on the key, and that Alice \& Bob's bit strings are identical, generally rely on the assumptions that i) the devices used to generate the key behave according to quantum theory, ii) the parties' locations are isolated to prevent unwanted information leakage, iii) Alice \& Bob get access to random numbers, iv) they can process classical information on trusted computers and v) their quantum devices are trusted and perfectly calibrated to carry out precisely the state and measurements foreseen by the protocol~\cite{Gisin02,Scarani09,Lo14, Pirandola2020}. 

\medskip 
When these assumptions are not met due to imperfections or simplifications in the QKD implementation, hacking becomes possible and compromises the security  of the key, see~\cite{Feihu2020} for a review of recent hacking experiments.
In order to reduce the vulnerability to these attacks, new QKD protocols relying on fewer assumptions are desirable. In device-independent QKD (DIQKD) especially, the assumption v) is removed. In other words, the state structure produced by the source, the underlying Hilbert space dimension and the operators describing the action of measurements apparatus are unknown a priori and their choice is even given to Eve. 

\medskip 
The higher security level of DIQKD comes at the price of a challenging implementation. In particular, DIQKD is entanglement-based~\cite{Ekert91} and requires high quality entangled states. Here, the entanglement quality is quantified by means of a non-local game: a high winning probability of the Bell-CHSH game~\cite{Clauser1969} for example, ensures that Alice \& Bob's state is closed to a two-qubit maximally entangled state in a device-independent way~\cite{Supic2020}, and therefore that the parties' measurement outcomes are unpredictable to any third party~\cite{Pironio2010}. Many entangled pairs are also required as the post-processing steps needed to distill an actual key from the outcomes of measurements on entangled states are bit consuming.  

\medskip 

Significant progress on the preparation of high quality entanglement using single trapped ions was recently reported experimentally~\cite{Stephenson2020}, and culminated with the first distribution of a device-independent key~\cite{Nadlinger2021}. Results have also been obtained to extend DIQKD over hundreds of meters with single atoms~\cite{Zhang2021}. A next step aims at implementing DIQKD with a purely photonic platform -- a platform where optical modes are entangled, manipulated and detected -- which is plausibly closer to what is expected for a commercial device. On the positive side, the Bell-CHSH game has already been properly implemented with a purely photonic platform using a photon pair source and photon counting techniques~\cite{Christensen13,Shalm15,Liu18,Shen18,Giustina15}. The reported winning probabilities are however very close to what can be obtained with classical strategies and are thus not sufficient to realize DIQKD\@. The main issue is the photon statistics emitted by sources used in these demonstrations which are very different from an ideal source producing two qubit states~\cite{CapraraVivoli15}. Another major problem is loss -- a significant fraction of photons are lost on the way from the source to the detectors~\cite{StateArtDIQKD2022}.
While the most advanced realization of DIQKD with photonic devices uses two-mode squeezing operations for creating polarization entanglement and photon detection combined with linear optical elements for polarization measurements~\cite{Liu2021}, a natural question is how to combine currently available photonic resources to facilitate the realization of DIQKD in this setting.

\medskip 

So far this question has not been addressed in an automated way. By combining Gaussian and non-Gaussian operations, researchers have imagined optical circuits that are capable of winning Bell games with a probability higher than classical strategies~\cite{Horne1989, Banaszek1999, Garcia2004}. Since the number of possible arrangements of optical elements grows exponentially with the number of operations considered, it is not clear however that all possible combinations of these operations have been considered. Therefore, simple configurations facilitating the implementation of DIQKD might have been missed. Recent developments of optical setups with integrated circuits also invite us to explore complex solutions with large numbers of modes and operations~\cite{Tanzilli2012,Pelucchi2021}. Furthermore, substantial theoretical efforts are devoted to the development of security proofs using the distribution of measurement results directly instead of bounding Eve's information from the winning probability of a Bell game obtained from this distribution~\cite{Sekatski2021,Woodhead2021,Brown2021,Tan2021,Kolodynski20,Bancal14,NietoSilleras14}.
The need to find optical circuits facilitating the implementation of DIQKD, the possibility of implementing complex optical circuits, the search for an optical setup producing exactly a given probability distribution of results and the constant evolution of DIQKD protocols push us to provide automated solutions to design optical experiments.

\medskip 

Machine learning~\cite{Lecun2015,Schmidhuber2015,Human2015,Silver2018} is becoming more and more useful in automation of problem-solving in quantum physics research~\cite{Biamonte2017, Dunjko2018, Carleo2019}. Inspired by Ref.~\cite{Melnikov20}, we introduce a technique combining reinforcement learning~\cite{Sutton2018}, optimisation algorithm~\cite{Nelder1965} and a custom efficient simulation of quantum optics experiments to design photonic setups maximizing a given function of the measurement statistics.
Applying the algorithm to DIQKD, it discovered new, unexpected experimental configurations leading to high key rates in both ideal and lossy cases. The relative simplicity of one of these settings together with its resistance to detector inefficiencies and noise, or the high key rate of a more advanced setting could be helpful for a first implementation of DIQKD and for future developments.

\bigskip

\paragraph{DIQKD protocol --} The protocol we consider uses a source that repeatedly creates a pair of entangled systems, half of each pair being sent to Alice, the other half to Bob. Each production of entangled systems defines a round. At each round, the parties measure their system according to a randomly chosen measurement. Alice in particular, can choose one out of two measurement settings labeled $\hat{A}_x$ with $x\in\{0,1\}$ and Bob has the choice between three measurement settings called $\hat{B}_y$ with $y\in\{0,1,2\}$. For each measurement input, one out of two possible outcomes is obtained that we label $A_x$ for Alice \& $B_y$ for Bob, with $\{A_x,B_y\}\in\{0,1\}$. The settings $x,y\in\{0,1\}$ are used in a CHSH game in which a round is won if the outcomes of Alice \& Bob are the same for the pair of settings $\{\hat{A}_0, \hat{B}_0\}$, $\{\hat{A}_0, \hat{B}_1\}$ and $\{\hat{A}_1, \hat{B}_0\}$ and different when the settings choice is $\{\hat{A}_1, \hat{B}_1\}$. The winning probability $\omega$ of the CHSH game is given by $(4+S)/8$ where the CHSH score $S$ is defined as 
 \begin{equation}
    S = \langle \hat{A}_0 \hat{B}_0 \rangle + \langle \hat{A}_0 \hat{B}_1 \rangle + \langle \hat{A}_1 \hat{B}_0 \rangle - \langle \hat{A}_1 \hat{B}_1 \rangle
\end{equation}
with 
\begin{equation}
    \langle \hat{A}_x \hat{B}_y \rangle = p(A_x = B_y |\hat{A}_x,\Hat{B}_y) - p( A_x \neq B_y |\hat{A}_x,\hat{B}_y).
\end{equation}
The setting $\hat{B}_2$ is ideally chosen to produce outcomes correlated with the results of $\hat{A}_0$. 

\medskip

At the end of these rounds, Alice (Bob) forms a raw key $\mathbf{A}$ ($\mathbf{B})$ from the results of her (his) measurements. The protocol then proceeds with an error correction step that allows Bob to reconstruct a copy of Alice's string $\mathbf{A}$. Using the choice of settings that Alice announces, Bob can estimate the Bell value $S$. In a final privacy amplification step, Alice \& Bob apply a randomness extractor to obtain the final secret key. In the asymptotic limit of a large number of rounds, the key generation rate when optimal one-way error correction and privacy amplification is used is given by~\cite{Devetak2005}
$
r = \entVN({\mathbf{A}}|E) - \entVN({\mathbf{A}}|{\mathbf{B}}), \label{keyrate}
$
with $\entVN$ the Von Neumann entropy. The first term, which quantifies Eve's uncertainty about the reference key ${\mathbf{A}}$, can be lower bounded from the function of a CHSH score~\cite{Pironio09}. When the protocol further includes a step where artificial noise is added to the measurement outcomes, Alice is instructed to generate a new raw key $\mathbf{A}'$ by flipping each of the bits of her initial raw key $\mathbf{A}$ independently with probability $p$ before the post-processing steps, Eve's uncertainty can increase depending on the value of $p$. In this case, the key rate is given by~\cite{Ho2020}
\begin{equation}
    r \leq 1 - I_p(S) - \entVN(\mathbf{A}'|\mathbf{B})
    \label{eq:keyrate}
\end{equation}
with
\begin{equation}
\begin{split}
    I_p(S) =\, &h \left(\frac{1+\sqrt{{(S/2)}^2-1}}{2} \right) \\
        - &h \left(\frac{1+\sqrt{1-p(1-p)(8-S^2)}}{2} \right),
\end{split}
\end{equation}
$h$ being the binary entropy. 

\bigskip
\paragraph{Photonic circuits under consideration --}  We consider an experiment involving $n$ bosonic modes initialized in the vacuum state. Their state is then manipulated by applying single-mode and two-mode operations on any mode or pair of modes in any order. $n-m$ of these modes are measured with non photon-number resolving detectors. The state preparation is finalized if the desired combination of measurement outcomes (click or no-click) is obtained on the measured modes. The remaining $m$ modes are split between Alice \& Bob. They apply on the received modes a local sequence of operations chosen from the same set which depends on their measurement settings. All the modes are finally detected by means of non-photon number resolving detectors, yielding one of the $2^m$ possible results. In the examples below, we explore circuits up to $\{m, n\} = \{2, 4\}$ to keep a reasonable implementation complexity. 

\medskip 

The set of possible unitary operations we consider is a fair representation of operations that are routinely used in quantum optics experiments. It includes single-mode squeezers, phase shifters and displacements for the single-mode operations, two-mode squeezers and beamsplitters for the two-mode operations. The use of photon detector is motivated by the need for non-Gaussian operations for obtaining statistics that cannot be reproduced by locally causal models and hence for producing a key device-independently.

\bigskip
\paragraph{Reference circuits --}   

\begin{figure}[htp]
    \centering
    \includegraphics[width=.45\textwidth]{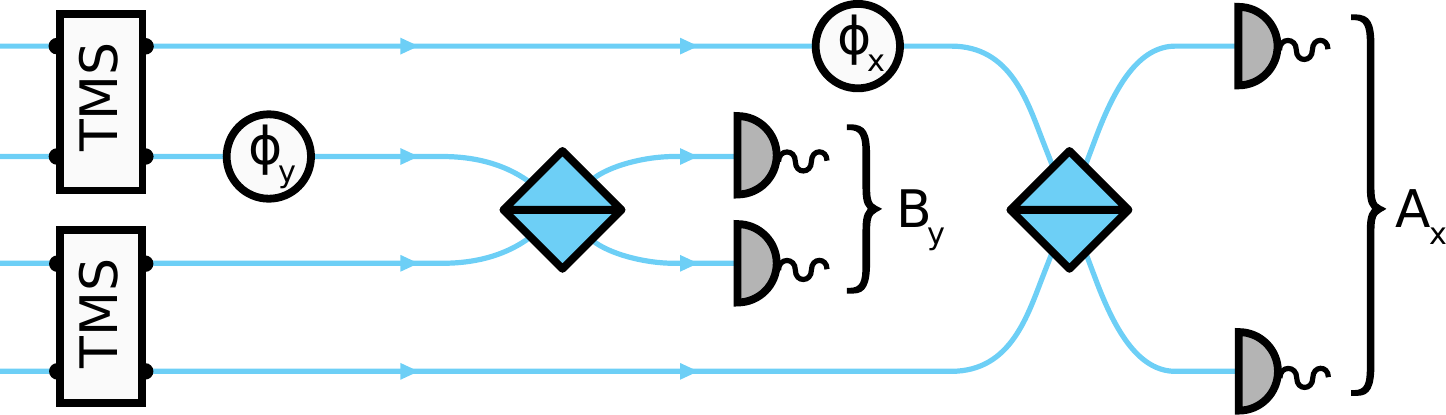}
    \caption{Most commonly used photonic experiment for realizing Bell tests which can naturally be envisioned for implementing device-independent quantum key distribution. Alice \& Bob receive each two modes, one from each of two two-mode squeezers (TMS rectangle) operating on the vacuum. The settings choice $\hat{A}_x$ ($\hat{B}_y$) are obtained by choosing the relative phase $\phi_x$ ($\phi_y$) of the two modes and the transmission of the beamsplitter (square) combining them. The measurements are finalized by placing two non-photon resolving detectors (grey half-circle) at the output of the beamsplitters. The outcomes $A_x$ and $B_y$ are obtained by binning the click and non-click events produced by the photon detectors.
    }\label{fig:pola}
\end{figure}

The most commonly used optical setup for realizing Bell
tests~\cite{Liu2021,Christensen13,Shalm15,Liu18,Shen18,Giustina15} uses a combination of two-mode squeezing operations acting on the vacuum for producing polarisation entangled photon pairs which are then measured with standard polarisation measurements. This setup is schematically depicted in Fig.~\ref{fig:pola}, where each of the $n=4$ bosonic modes (spatial and polarization) is represented by a separate line. The settings are chosen by varying the relative phase between the modes and the transmission/reflection of the beamsplitter. By optimizing the squeezing parameters, the setting choice, the amount of noise in the noisy post-processing and by binning the measurement results appropriately, the formula~\eqref{eq:keyrate} yields a key rate of $~\sim 0.2522$ in the ideal case, that is without noise and loss~\cite{Ho2020}. When considering detectors with non-unit efficiencies, the key rate decreases as shown in Fig.~\ref{fig:key_rates_comparaison}, see blue dashed line. The critical detection efficiency, that is the minimum detection efficiency that is needed to generate a positive key rate is of $82.6\%$~\cite{Ho2020}. This serves as a reference to benchmark the performance of any alternative circuits enabling DIQKD\@.

\bigskip
\paragraph{Simulating quantum optical circuits --} To explore the set of possible photonic circuits and quantify their performance, we need to be able to simulate them and compute the measurement statistics.  Furthermore, it is desirable that such simulation is efficient as it will set a bottleneck on the performance of automated circuit design.

\medskip 

To represent the state prepared by an optical circuit and deduce the measurement statistics, we use the first and second moments of quadrature operators. Formally, 
if $a_i$, $a_i^\dagger$ are the bosonic operators for the mode $i=\{1,\dotsc,n\}$, the corresponding dimensionless quadrature operators are given by $\hat x_i = \frac{a_i^\dagger + a_i}{2}$ and $ \hat p_i = \ii \frac{a_i^\dagger - a_i}{2}$ with $[\hat x_i,\hat p_i]=\frac{\ii}{2}$. We collect these $2n$ operators in a vector
\begin{equation}
    \mathbf{q} = (\hat x_1,\hat p_1,\ldots,\hat x_n,\hat p_n)
\end{equation}
and label the $i$th component of this vector $q_i$. The displacement vector $\bmu$ and the covariance matrix $\Sigma$ associated to a given state are defined as the expected values
\begin{align}
    \mu_i &= \mean{q_i}, \\
    \Sigma_{ij} &= \frac{1}{2}\mean{q_i q_j + q_j q_i} - \mu_i \mu_j.
\end{align}  
on the state.
$\bmu$ and $\Sigma$ give a faithful representation of any $n$-mode Gaussian states in terms of $2n^2 +3 n$ real parameters, see~\cite{Weedbrook2012,Adesso2014, Brask2021} and the Appendix~A.
In the optical setups that we consider, most of currently used operations are Gaussian. As long as such operations only are applied on the initial vacuum, the state prepared by the circuit remains Gaussian and can be represented by the displacement vector and the covariance matrix.

\medskip 

The only exception that we consider which is capable of producing non-Gaussianity is the photon detection. Nevertheless, as we show in the Appendix A, if one starts with a $n$-mode Gaussian state and measures out one mode with a single photon detector, the state of the remaining $n-1$ modes conditioned on the no-click outcome is also Gaussian. So is the state of the remaining  modes when one of the $n$ modes is traced out. It follows that the state of the $(n-1)$-modes conditioned on the click outcome can be written as a difference between two Gaussian states. Hence, the state resulting from a preparation circuit with a single mode used as a conditioning measurement can be fully described by two pairs of displacement vectors and covariance matrices. For each additional heralding operation, the numbers of parameters required to describe the state has to be doubled. Nevertheless, if the number of modes used for heralding remains low, we obtain a memory efficient exact representation of the state associated with a circuit. This is precisely our regime of interest, since we want a reasonable heralding rate to end up with feasible proposals.



\bigskip
\paragraph{Automated design of quantum optics experiments --} The automated design of quantum optics experiments is based on reinforcement learning -- a machine learning paradigm in which an agent is interacting with an environment and learns a task by trials and errors. The agent is a routine which specifies the order with which operations are placed on the different modes. The environment efficiently models the series of operations proposed by the agent in order to deduce the measurement statistics and set the parameters of chosen operations to optimize the key rate. The maximal key rate computed by the environment is fed back to the agent as a reward.

\medskip

The task of the agent is to invent photonic circuits suitable for DIQKD\@. It learns to do so by repeatedly interacting with a virtual optical circuit inside an episode until a stop condition is met. Specifically, at the beginning of each episode $e$, the agent perceives a state $s_1(e)$ which is a representation of the (empty) optical circuit at the first step $k=1$. After a deliberation phase, the agent places an optical element or a series of optical elements corresponding to an action $a_1(e)$ on the bosonic modes. This produces a new circuit which is analyzed by the environment. The agent then receives back the new state of the circuit $s_2(e)$ together with the associated reward $r_2(e)$ which  can be adapted depending on the property of circuits that is desired. An interaction step starts again and the end of the episode is reached when a given circuit depth is achieved. The agent learns from past experiences $\{s_i(k),r_i(k)\}$ by updating the policy behind the deliberation process.

\medskip 

The task of the environment is to simulate the optical circuits proposed by the agent and optimise its parameters in order to compute the reward associated to the circuit. To simulate the circuit and deduce the measurement statistics, we use the memory efficient representation discussed earlier based on the displacement vectors and covariance matrices. The package that we developed is made available \textsc{QuantumOpticalCircuits.jl}~\cite{Valcarce2021} (written in \textsc{Julia}~\cite{Bezanson2017} to make it fast and extensible). All of the unitary operations and photon detections described in the previous section are included. 
To find the parameters of circuits chosen by the agent leading to the highest key rates according to the bound given in~\eqref{eq:keyrate}, we used the Nelder-Mead algorithm, a suitable algorithm for the optimization of multidimensional non-linear objective functions which does not require an analytical or numerical gradient to be supplied.
The optimization is performed in the ideal case, i.e.\@ with unit detector efficiency. Each time the optimisation is called, a new set of parameters is chosen as a starting point to avoid local minima. To get access to the critical detection efficiency, the optimal parameters are computed by first considering detectors with unit efficiencies. The efficiency is then decreased and a new parameters optimisation is performed, starting from the ones resulting in the best key rate at the previous step.
The process starts again with a smaller detection efficiency until the key rate drops below a certain threshold.


\begin{figure}[htp]
    \centering
    \includegraphics[width=.45\textwidth]{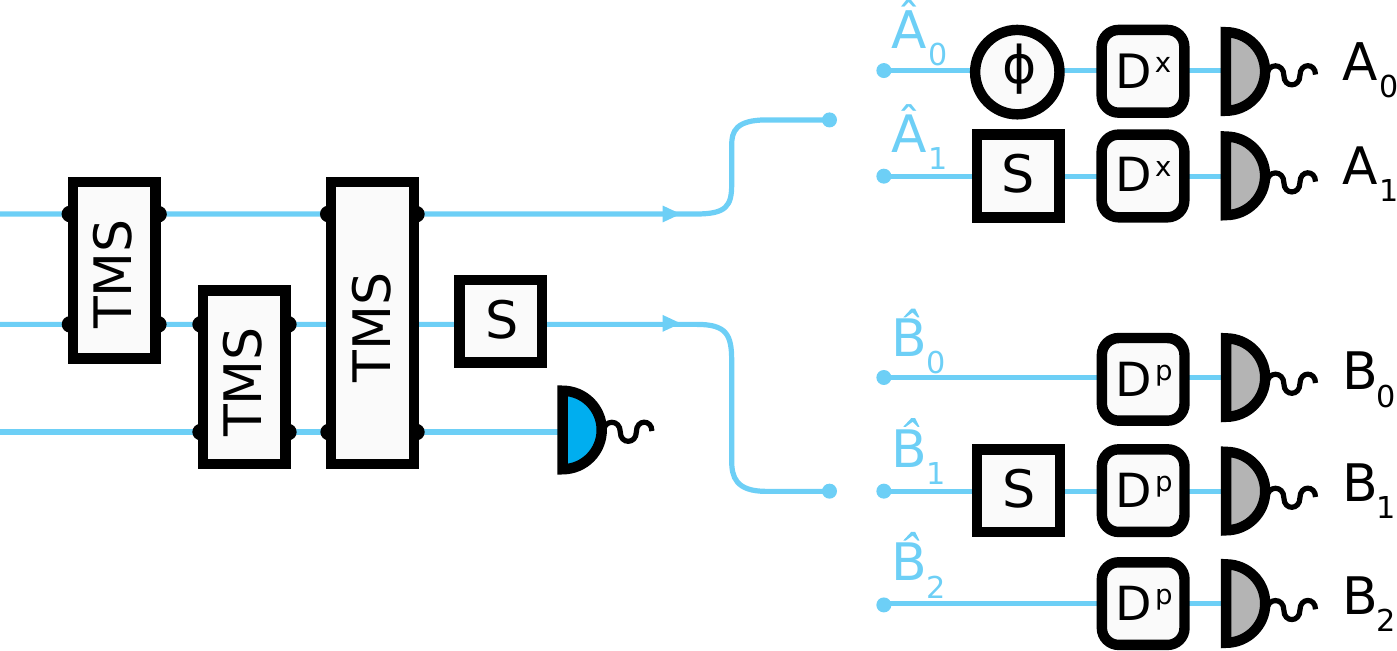}
    \caption{%
   Setup offering the highest key rate in the absence of loss and noise. The state is prepared by a first succession of three two-mode squeezed states (TMS) operating on the pair of modes $\{1,2\}$, $\{2,3\}$ and $\{1,3\}$ respectively. They are then followed by a single mode squeezers (S) on mode $2$. The state of modes $1$ and $2$ conditioned on a click at the photon detector on mode $3$ is sent to Alice \& Bob. For each of his measurement settings $y=\{0,1,2\}$, Bob performs a displacement operation in the $p$ direction (D$^p$). In the case $y=2$ a single-mode squeezer is applied before the displacement operation. On Alice's side, the first measurement setting is obtained by applying a phase shifter on the mode she received before applying a displacement operation along $x$ (D$^x$). For the second setting, a single-mode squeezer replaces the phase shifter and is followed by another displacement in $x$. The optimal parameters of each operation, for different losses value, are given in Tab.~\ref{tab:eff_1}.
   }\label{fig:setup_eff_1}
\end{figure}

\bigskip
\paragraph{Choice of the policy --} Transition in the environment, i.e.\@ the change in state and reward from an action, is the result of a heuristic numerical optimisation. In such a case, it is natural to use model-free reinforcement learning, learning purely from trial-and-errors, not trying to construct a model of that transition~\cite{Sutton2018}.
Furthermore, infinitely many states can be observed and learning which action performs best for each state is computationally impossible. Instead, we use policy gradients method, which aim at learning directly a stochastic policy mapping states to actions~\cite{Weng2018}.
Finally, the sampling cost, i.e.\@ the cost to simulate an episode, is high from the numerical optimisation. In this scope, we used the proximal policy optimisation (PPO) algorithm~\cite{Schulman2017}, a model-free policy gradient algorithm known to be sample efficient compared to similar reinforcement learning algorithms, e.g. TRPO~\cite{Schulman2015}. 
The details of this algorithm are available in the Appendix B.

\bigskip
\paragraph{Results --} In the first step, we define a reward which favors a high key rate in the ideal case, that is, in the absence of loss and noise. The setup found by the agent after a few training steps is depicted in Fig.~\ref{fig:setup_eff_1}. It involves three modes, one mode serving as a heralding after a series of two-mode and single-mode squeezed operations. This setup yields a key rate of $\sim 0.914$, much higher than the reference circuit which yields a key rate of $\sim 0.252$. The resistance to loss of this unexpected setup is characterized by optimizing the key rate as a function of the detection efficiency $\eta$ (the detectors of Alice \& Bob and the one used for the heralding are assumed to have the same efficiency). The result is shown in~Fig.\ref{fig:key_rates_comparaison}, see orange dashed line. We see that the setup proposed in Fig.~\ref{fig:setup_eff_1} provides a higher key rate than the reference circuit for detection efficiencies larger than $\sim 87.5\%$. 

\begin{figure}[htp]
    \centering
    \includegraphics[width=.45\textwidth]{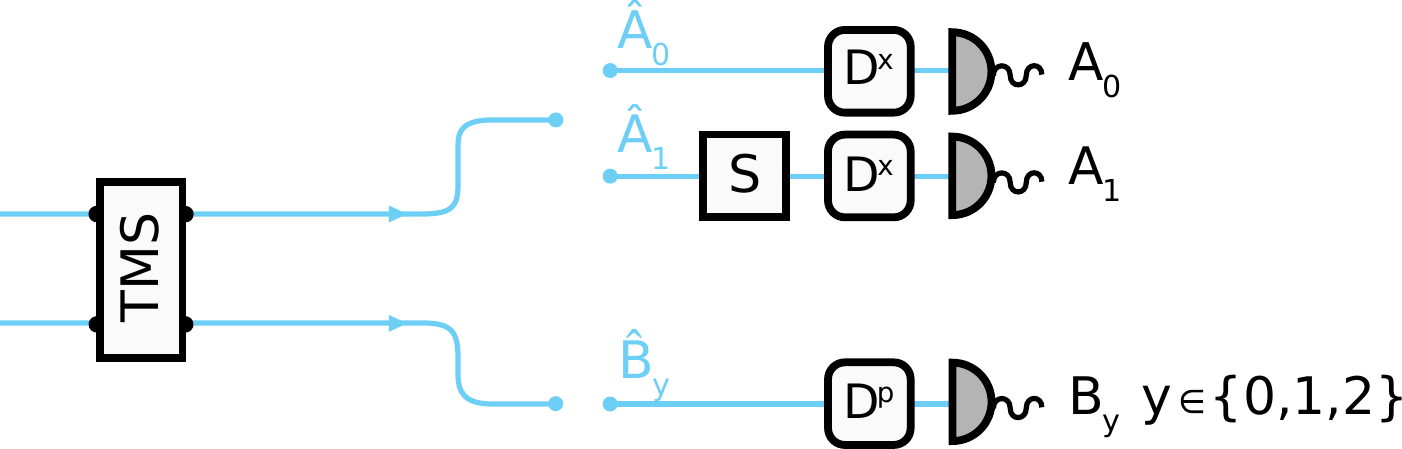}
    \caption{Setup with the highest tolerance to detector inefficiency when considering key rates greater than $10^{-4}$. The preparation step is made with a single two-mode squeezer (TMS) operating on vacuum. Alice's settings are set by either a displacement for $\hat{A}_0$ or a single mode squeezer followed by a displacement for $\hat{A}_1$. Bob's settings correspond to displacement operations with different amplitudes for each input.
    }\label{fig:best_setup}
\end{figure}
\medskip

In the second step, we adapt the reward to favor loss tolerant circuits. Concretely, we look for circuits with a minimum detection efficiency for achieving a key rate of at least $10^{-4}$. Interestingly, the most interesting setup was found for $n=2$. As shown in Fig.~\ref{fig:best_setup}, this circuit uses a single two-mode squeezed operation applied on the vacuum. While Bob's settings correspond to displacements with three different amplitudes, Alice's settings consist either in a single displacement operation or in a combination of single-mode squeezing and displacement operations. As shown in Fig.~\ref{fig:key_rates_comparaison} (green full line), this circuit is able to produce a key rate of $\sim 10^{-8}$ for a detection efficiency of $82.45\%$, while for the reference circuit the same rate requires a detection efficiency of $83.25\%$. Moreover, depending on the detection efficiency, the key rate can be up to 2 orders of magnitude larger. Finally, we also compare the resistance to some noise. In particular, when including dark counts with a probability of $10^{-3}$, the reference circuit yields a key rate of $10^{-4}$ provided the detection efficiency is of $\sim 91.1\%$ while the proposed circuit needs a detection efficiency of $\sim 86.1\%$ for achieving the same key rate. Note that the parameters of circuits presented in Figs.~\ref{fig:setup_eff_1} and~\ref{fig:best_setup} are given in Appendix C.  

\begin{figure}[htp]
    \centering
    \includegraphics[width=.4\textwidth]{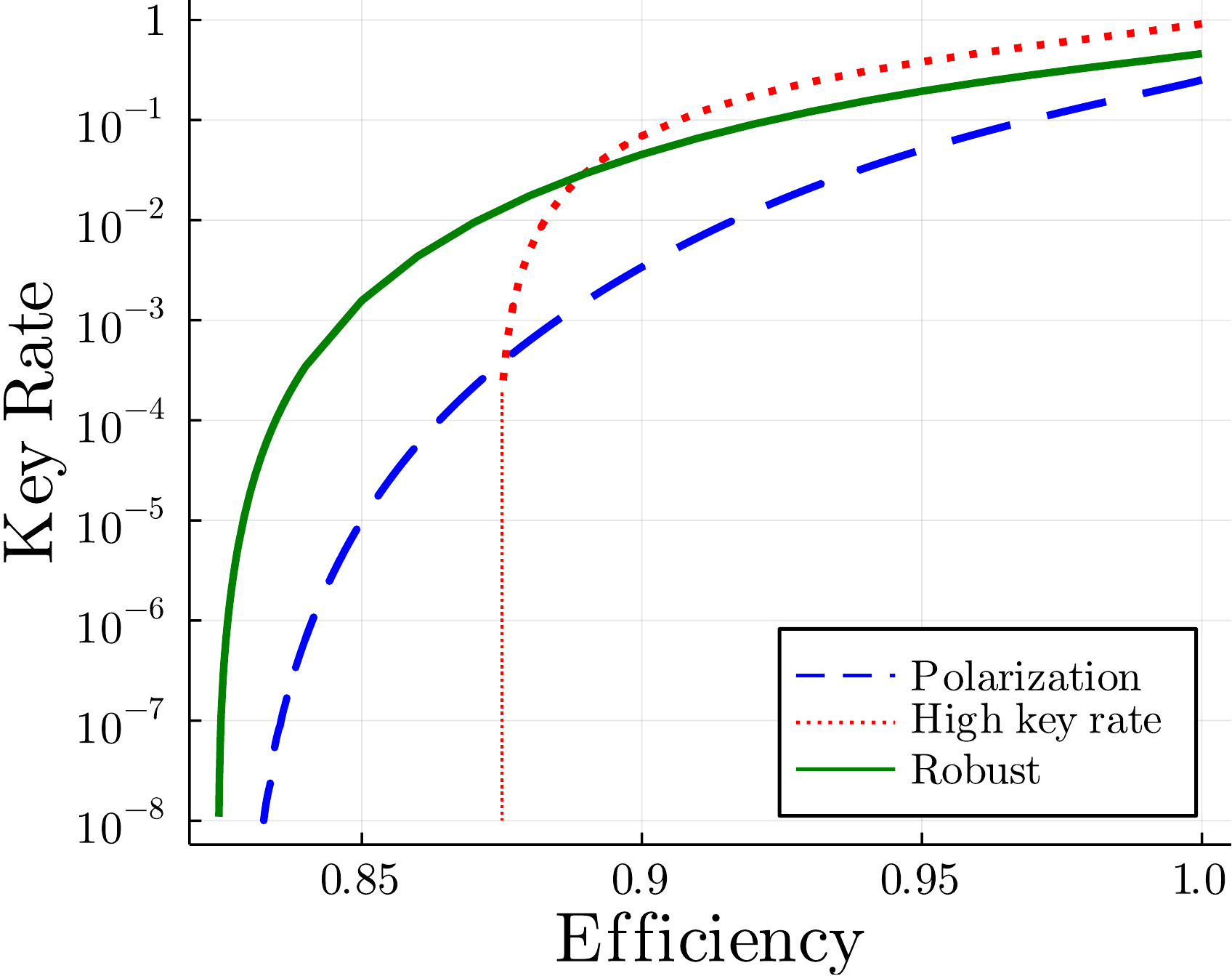}
    \caption{Key rate as a function of the detector efficiency. The blue dashed line is associated with the reference circuit, i.e.\@ the standard case with four modes coupled two by two by squeezing operations (Fig.~\ref{fig:pola}). The orange dotted line corresponds to the circuit found when high key rates are favored for unit efficiency detections (Fig.~\ref{fig:setup_eff_1}). The green solid line is associated to the circuit found when favoring the resistance to non-unit detection efficiency (Fig.~\ref{fig:best_setup}).}\label{fig:key_rates_comparaison}
\end{figure}

\bigskip
\paragraph{Conclusion --}

We presented an algorithm combining an efficient modeling of Gaussian and heralded processes, an optimisation and a reinforcement learning agent to automate the process of finding optical circuits producing desired statistics. We showed its potential usefulness for implementing DIQKD for both mid and long term goals aiming respectively to facilitate a first photonic realization and to deliver high key rates. We do expect the proposed algorithm to remain useful in case new protocols and security proofs are proposed as the formula of the key rate that was used to guide the efforts of the agent can be updated to include future developments.

\bigskip
\paragraph{Acknowledgment --}
The author would like to thank Jean-Daniel Bancal, Jean Etesse, Anthony Martin, Ernest Tan, M\"arta Tschudin, Ramona Wolf, Julian Zivy for fruitful discussions and Dowinson Nguyen for the illustrations. We acknowledge funding by the Institut de Physique Théorique (IPhT), Commissariat à l’Energie Atomique et aux Energies Alternatives (CEA) and by a French national quantum initiative managed by Agence Nationale de la Recherche in the framework of France 2030 with the reference ANR-22-PETQ-0009\@.

\bibliography{references}

\newpage

\begin{widetext}



%

\section{Appendix A. Modeling the photonic circuits}

In this appendix we present a concise and self-contained derivation of how we model the photonic circuits discussed in the main text with a low number of parameters. We start by giving a short summary containing all the formulas required to use this representation. The derivations are presented afterwards.

\subsection{Summary}
 Consider $n$ bosonic modes and the associated quadrature operators  $ \hat x_i = \frac{ a_i^\dagger +  a_i}{2}$ and $ \hat p_i = \ii \frac{ a_i^\dagger -  a_i}{2}$ that we collected in a vector  $\mathbf{q} = ( \hat x_1,  \hat p_1,\ldots,  \hat x_n, \hat p_n) = (q_1,q_2,\dots , q_{2n-1}, q_{2n})$.
Any Gaussian $n$-mode state $\rho$ can be faithfully represented by the first
two moments  $(\bmu,\Sigma)$ of the quadrature operators on the state $\mu_i = \mean{q_i} = \tr \rho \, q_i$ and
    $\Sigma_{ij} =\frac{1}{2}\mean{q_i q_j + q_j q_i} - \mu_i \mu_j$.
Here $\bmu$ is the displacement (column) vector ($2n$ real parameters) and $\Sigma=\Sigma^T$ is the covariance matrix ($2n^2 + n$ real parameters). In particular, for the vacuum state one finds
\begin{equation}
\text{Vacuum state} \, \ketbra{0}^{\otimes n} \quad : \quad (\bmu, \Sigma) = \left(\vec{0}, \frac{1}{4} \mathds{1}\right).
\end{equation}

A Gaussian operation $\text{T}$ maps Gaussian states to Gaussian states. It can  be represented by means of a pair $(\bm d,M)$ with a $2n$ real column vector $\bm d$ and a symplectic $2n \times 2n $ matrix $M$. When acting on a state the Gaussian operation transforms the displacement vector and the covariance matrix as 
\begin{equation}
\text{T}: (\bmu,\Sigma) \mapsto (M \bmu + \bm d ,M \Sigma M^T)
\end{equation} In the Tables~\ref{tab:single mode} and \ref{tab:two mode} we give the representation of all single-mode and two-mode Gaussian transformations in term of $(\bm d,M)$ as well as the  corresponding unitary representation $\text{T}: \ket{\psi}\mapsto U\ket{\psi}$. We use the notation $\text{c}_\theta=\cos(\theta), \text{s}_\theta=\sin(\theta), \text{C}_r= \cosh(r)$, $\text{S}_r = \sinh(r)$.
\begin{table}[h]
    \centering
    \begin{tabular}{|c|c|c |c |c|}
        \hline Operation  & $U$ &$\bm d$ &$M$ & Parameters\\ \hline &&&& 
        \\
          Displacement (D)& $e^{\alpha a_i^\dag - \alpha^* a_i}$ &$(\dots 0,\Re{\alpha}, \Im{\alpha},0 \dots )^T$ & $\mathds{1}_{2n}$ & $\alpha\in \mathbb{C}$\\
          Phase shifter ($\Phi$)& $e^{-\ii \theta a_i^\dag a_i}$ &$\vec 0$ & $\left(\begin{array}{cc}
    \text{c}_\theta & \text{s}_\theta\\
    -\text{s}_\theta & \text{c}_\theta
    \end{array}\right)\oplus \mathds{1}_{2n-2}$ & $\theta\in\mathbb{R}$ \\
    Single-mode squeezer (S)& $e^{\frac{1}{2}(z^*a_i^2 - za_i^{\dag 2})}$&$\vec{0}$ & $\left(\begin{array}{cc}
   \text{C}_r - \text{S}_r \text{c}_\theta &- \text{S}_r \text{s}_\theta \\
    -\text{S}_r \text{s}_\theta & \text{C}_r +\text{S}_r \text{c}_\theta
    \end{array}\right) \oplus \mathds{1}_{2n-2}$ & $z=r e^{\ii \theta}  \in \mathbb{C}$  \\
    \hline
    \end{tabular}
    \caption{%
        Single-mode Gaussian operations. When the operation acts on the $i$th mode,  the nonzero components of $\bm d$ and the nontrivial block of the matrix $M$ appear at positions $2i-1$ and $2i$. For the displacement we also consider the particular cases denoted $\text{D}^{\hat x}$ for $\Im\{\alpha\}=0$ and $\text{D}^{\hat p}$ for $\Re\{\alpha\}=0$.
    }\label{tab:single mode}
\end{table}

\begin{table}[h]
    \centering
    \begin{tabular}{|c|c |c |c |c|}
        \hline Operation  & $U$ & $\bm d$ &$M$ &  Parameters
        \\ 
        \hline &&&& 
        \\
          Beamsplitter (BS)&  $e^{\theta(a_i^\dag a_j - a_i a_j^\dag)}$ &$\quad \vec {0} \quad$ & $\left(\begin{array}{cccc}
    \text{c}_\theta & 0 & \text{s}_\theta & 0 
    \\
    0 & \text{c}_\theta & 0 & \text{s}_\theta \\
    -\text{s}_\theta & 0 & \text{c}_\theta & 0 \\
    0 & -\text{s}_\theta & 0 & \text{c}_\theta 
    \end{array}\right) \oplus \mathds{1}_{2n-4}$ &$\theta\in \mathbb{R}$ 
    \\
    Two-mode squeezer (TMS)& $e^{z^*a_i a_j - z a_i^\dag a_j^\dag} $ & $\vec{0}$ &$\left(\begin{array}{cccc}
    \text{C}_r & 0 & -\text{S}_r \text{c}_{\theta} & -\text{S}_r \text{s}_{\theta} \\
    0 & \text{C}_r & -\text{S}_r \text{s}_{\theta} & \text{S}_r \text{c}_{\theta} \\
   -\text{S}_r \text{c}_{\theta} & -\text{S}_r \text{s}_{\theta} & \text{C}_r & 0 \\
    -\text{S}_r \text{s}_{\theta} & \text{S}_r \text{c}_{\theta} & 0 & \text{C}_r
    \end{array}\right) \oplus \mathds{1}_{2n-4}$
    & $z = re^{\ii \theta}\in \mathbb{C}$  \\
    \hline
    \end{tabular}
    \caption{Two-modes Gaussian operations. When the operation acts on modes $i$ and $j$ the nontrivial block of the matrix $M$ appears at positions $2i-1,2i,2j-1$ and $2j$}\label{tab:two mode}
\end{table}

\medskip 

Next we consider the transformations of the state resulting form measuring out the mode $i$ with a single photon detector of efficiency $\eta$. Both of the states $\rho_{\lnot i}$ and $\rho_{\circ i}$ of the remaining $(n-1)$ modes resulting respectively from discarding the measurement outcome (or tracing out the mode i, $\rho_{\lnot i}$) and conditioning on the no-click outcome ($\rho_{\circ i})$ are Gaussian. The displacement vectors and  quadrature moments of the resulting states are given in the Table.~\ref{tab:conditioning}. To express the state we use the transformation $\text{TR}_i$ which simply drops the components of a vector or rows at columns of a matrix at positions $2i-1,2i$
\begin{equation}
    \text{TR}_i[\bmu] \equiv  (\dots, \mu_{2i-2},\mu_{2i+1},\dots)\qquad 
    \text{TR}_i[\Sigma] \equiv
    \left(\begin{array}{cc|cc}
    \ddots & \vdots & \vdots &\\
    \dots& \Sigma_{2i-2,2i-2} & \Sigma_{2i-2,2i+1} & \dots\\
    \hline
    \dots& \Sigma_{2i+1,2i-2} & \Sigma_{2i+1,2i+1} &\dots  \\
    & \vdots& \vdots& \ddots
    \end{array}\right), 
\label{eq: TR}
\end{equation}
and the matrix $F$ given by 
\begin{equation}
F= \left(\begin{array}{cc}
    \frac{4\eta}{2-\eta} & 0 \\
    0 & \frac{4\eta}{2-\eta}
    \end{array}\right)_{2i-1,2i} \oplus 0_{2n-2}.
\label{eq: F matrix}    
\end{equation}
\begin{table}[h]
    \centering
    \begin{tabular}{|c|c |c |c|}
        \hline Transformation & density matrix & displacement vector & covariance matrix \\ \hline
          Tracing out the mode $i$ (discarding the outcome) & $\rho_{\lnot i}$ & $\bmu_{\lnot i}\equiv \text{TR}_i(\bmu)$ & $\Sigma_{\lnot i}\equiv \text{TR}_i(\Sigma)$ \\
    Conditioning to no-click outcome on the mode $i$ & $\rho_{\circ i}$ & $\bmu_{\circ i}\equiv \text{TR}_i[(\Sigma^{-1}+F)^{-1} \Sigma^{-1} \bmu]$ & $\Sigma_{\circ i}\equiv \text{TR}_i[(\Sigma^{-1} + F)^{-1}]$\\
    Conditioning to click outcome on the mode $i$  & $\rho_{\bullet i}$ & $\cross$ & $\cross$\\
    \hline
    \end{tabular}
    \caption{%
        The possible states resulting from a $n$-mode Gaussian state $\rho \simeq (\bmu,\Sigma)$ after measuring out the mode $i$ with a single photon detector with efficiency $\eta$. The transformation $\text{TR}_i$ defined in Eq.~\eqref{eq: TR} simply removes the components of a vector or rows and columns of a matrix at positions $2 i-1$ and $2i$. The matrix $F$  defined in Eq.~\eqref{eq: F matrix} depends on the detector efficiency. The state $\rho_{\bullet i}$ is not Gaussian but can be decomposed as a difference of two Gaussian states, see Eq.~\eqref{eq: rho cond click}.
    }\label{tab:conditioning}
\end{table}

\noindent The probabilities to observe the no-click ($p_{\circ i}$) and the click ($p_{\bullet i}$) outcomes are given by 
\begin{equation}
p_{\circ i} =\frac{2}{2-\eta}\sqrt{\frac{(\det \Sigma)^{-1}}{ \det(\Sigma^{-1}+F)}}
    e^{-\frac{1}{2}{\bm \mu }^T\left( \Sigma^{-1} - \Sigma^{-1} (\Sigma^{-1}+F)^{-1} \Sigma^{-1} \right){\bm \mu }} \quad \text{and} \quad  p_{\bullet i} = 1-p_{\circ i}.
\end{equation}
Importantly, the state $\rho_{\bullet i}$ conditional on the click outcome is non-Gaussian, but can be expressed as a difference  of two Gaussian states
\begin{equation}
    \label{eq: rho cond click}
    \rho_{\bullet i} = \frac{\rho_{\lnot i} - p_{\circ i} \rho_{\circ i}}{1 - p_{\circ i}}\quad\simeq\quad
        \left\{
        \begin{array}{c} \frac{1}{1-p_{\circ i}},(\bmu_{\lnot i},\Sigma_{\lnot i})\\
            -\frac{p_{\circ i}}{1-p_{\circ i}},(\bmu_{\circ i},\Sigma_{\circ i})
        \end{array}
        \right\}.
\end{equation}
For any subsequent heralding, the number of terms in the sum is doubled. Generally, we are thus interested in states $\rho = \sum_k w_k \rho_k$ that can be represented as a quasi-mixture ($\rho \simeq \{ w_k, (\bmu_k,\Sigma_k)\}$) of Gaussian state $\rho_k$, where the weights $w_k$ can be negative. In total such a representation requires $2^{n-m} (2m^2 +3 m+1)$ real parameters, where the number of modes used for heralding $n-m$ is kept low for the setups of interest.

\medskip

Finally, we give a compact formula that computes the measurement statistics. When $m$ modes are measured with single photon detectors, the outcomes are labeled by a bitstring $\bm k$ of length $m$, where each bit specifies if the detector on the mode $i$ clicks ($k_i=1$) or not $(k_i=0)$. For a Gaussian state described by $(\bmu,\Sigma)$ the probability to observe the outcome $\bm k$ is given by 
\begin{align}
    &\text{Prob}(\bm k) = \sum_{\bm \ell \in \mathds{S}_{\bm k}} (-1)^{\bm k\cdot\bm \ell}  \mean{\hat{o}_{\bm \ell}}
    \quad \text{with} \\
   \mean{\hat{o}_{\bm \ell}} &=\left(\frac{2}{2-\eta}\right)^{|\bm \ell|} \frac{\exp\left(-\frac{1}{2}\bm \mu^T (\Sigma^{-1} - \Sigma^{-1}(\Sigma^{-1}+O_{\bm \ell})^{-1}\Sigma^{-1} )\bm \mu \right)}{\sqrt{\det(\Sigma)\det(\Sigma^{-1}+O_{\bm \ell})}}, \quad
   O_{\bm \ell} = \frac{4\eta}{2-\eta} \bigoplus_{i=1}^m \begin{pmatrix}
\ell_i  & 0\\
0& \ell_i 
\end{pmatrix}.
\end{align}
Here $\mathds{S}_{\bm k}$ is the set of all bitstrings $\bm \ell$ of length $m$ whose components $\ell_i$ are fixed to $1$ for all $i$ such that $k_i=0$ (a set of size $2^{|\bm k|}$),  and $|\bm \ell| = \sum_{i=1}^{m} \ell_i$ is the Hamming weight of a bitstring.

\subsection{Wigner representation}

Let us first consider a single bosonic mode associated with creation and annihilation operators $a^\dag$ and $ a$. The Dirac delta operator, defined as
\begin{equation}
\delta(a- \alpha) = \frac{1}{\pi^2} \int d^2 \beta \,  e^{( a^\dag - \alpha^*)\beta -( a -\alpha)\beta^*}
\end{equation}
where $\alpha$ and $\beta$ are complex numbers, possesses various properties and in particular
\begin{equation}
\label{quasi_proba}
  \pi \int d^2 \alpha\,  \delta( a-\alpha) \tr \rho \, \delta( a-\alpha) =\rho
\end{equation}
for any density operator $\rho$, see Ref.\cite{Vogel2006} Section $4$. This suggests that $\delta( a- \alpha)$ can be used for representing density operators with the quasi-probability distribution 
\begin{equation}
\label{eq: Wigner def}
    W_\rho(\alpha) = \tr\,  \rho \, \delta( a-\alpha)
\end{equation}
satisfying $-2\pi^{-1} \leq W_\rho(\alpha) \leq 2\pi^{-1}$~\cite{Vogel2006}. This representation -- the Wigner representation -- can be extended to the multi-mode case. Consider $n$ bosonic modes with the operator $a_i,a_i^\dag$ associated to the mode $i=\{1,n\}$. The Wigner representation of an $n$-mode state $\rho$ is defined by the following extension of the mono-mode case
\begin{equation}\label{eq: Wigner multi}
        W_\rho({\bm \alpha}) = \tr \, \rho \, \bigotimes_{i=1}^n \delta(a_i-\alpha_i)
\end{equation}
where $\balpha=\{\alpha_1, \hdots, \alpha_n\}^T \in \mathds{C}^n$.
As before, the link between the state $\rho$ of the $n$ modes and the Wigner function is given by 
\begin{equation}
 \rho = \pi^n \int d^2{\bm \alpha}\, W_\rho({\bm \alpha})\,   \bigotimes_{i=1}^n \delta( a_i-\alpha_i).
\end{equation}

\subsection{Wigner representation of the vacuum state}
A $n$-mode state $\rho$ is called Gaussian if its Wigner function is Gaussian, i.e.\@ equal to probability density function $\t{N}({\bm \alpha}; {\bm \mu},\Sigma)$ of a multivariate normal distribution~\cite{Adesso2014,Weedbrook2012}
\begin{equation}
    \label{eq: Wigner}
   W_\rho(\balpha) = \t{N}({\bm \alpha}; {\bm \mu},\Sigma) = \frac{\exp(-\frac{1}{2}(\tilde{\balpha}-\bmu)^T\Sigma^{-1}(\tilde{\balpha}-\bmu))}{\sqrt{\det(2\pi\Sigma)}}.
\end{equation}
$\tilde{\balpha}$ is the $\mathds{R}^{2n}$ vector constructed from $\balpha$ in the following way
\begin{equation}
      \tilde{\balpha} = \{\t{Re}{(\alpha_1)}, \t{Im}{(\alpha_1)}, \hdots, \t{Re}{(\alpha_n)}, \t{Im}{(\alpha_n})\}^T.
\end{equation}
$\bmu$ -- the displacement (column) vector -- and $\Sigma$ -- the covariance matrix. Together they are given by $2n^2 +3 n$ real parameters with
\begin{align}
    \mu_i &= \mean{q_i}, \qquad \\
    \Sigma_{ij} &= \frac{1}{2}\mean{q_i q_j + q_j q_i} - \mu_i \mu_j.
\end{align}
with $q_i$ the $i$th component of the vector
    $\mathbf{q} = (\hat x_1, \hat p_1,\ldots, \hat x_n, \hat p_n)$ composed of  the dimensionless quadrature operators  $\hat  x_i = \frac{ a_i^\dagger +  a_i}{2}$ and $\hat p_i = \ii \frac{ a_i^\dagger -  a_i}{2}$ which satisfy $[\hat  x_i,  \hat p_i]=\frac{\ii}{2}$.

\medskip 

A single mode vacuum state, $\ket{0}$, is an example of  a Gaussian state. Its Wigner function is characterized by a zero displacement vector $\bmu=\vec 0_2$ (as $\bra{0}  \hat x \ket{0}=\bra{0}  \hat p \ket{0}=0$) and a  covariance matrix proportional to identity $\Sigma =\frac{1}{4}\mathds{1}_{2}$ (as $\bra{0} {\hat  x}^2 \ket{0}=\bra{0} {\hat  p}^2 \ket{0}= \frac{1}{4}$ and $\bra{0} \{ \hat x, \hat  p\} \ket{0}= \bra{0} \frac{\ii}{4}(a a^\dag -{a}{a}^\dag ) \ket{0}= 0$).  
Extending this to $n$ modes, we get that the Wigner function of the $n$-mode vacuum state is given by~\eqref{eq: Wigner} with 
\begin{align}
    \bmu &= \vec{0}_{2n},\ \\
    \Sigma &= \frac{1}{4}\mathds{1}_{2n},
\end{align}
where the subscript specifies the size of the objects.

\subsection{Gaussian operations} 
Among possible operations on bosonic systems, Gaussian operations are those mapping Gaussian states to Gaussian states. They are thus fully characterized by their effect on the displacement vector and covariance matrix~\cite{Adesso2014}
\begin{equation}
    \bmu \mapsto M \bmu+\bm d \quad \text{and} \quad \Sigma \mapsto    M\Sigma M^{T}
\end{equation}
where, for a $n$-mode system, $\bm d\in \mathbb{R}^{2n}$ and $M$ is a symplectic matrix, i.e.\@ a $2n\times2n$ matrix satisfying
\begin{equation}
    M^{T}\Omega M = \Omega \qquad \text{with}\qquad 
    \Omega = \mathds{1}_n \otimes \left(\begin{array}{cc}
            0 & 1 \\
            -1 & 0
            \end{array}\right).
\end{equation}
Note that a single mode Gaussian operation characterized by $M$ and $\bm d$ acting on the mode $i$ of a $n$-mode state has a vector $\bm d = (\dots,0, d_{2i-1},d_{2i},0\dots)^T$ with all elements that are zero except at positions $2i-1$ and $2 i$, and  a symplectic matrix $M= \mathds{1}_{2(i-1)} \oplus M' \oplus \mathds{1}_{2(n-i)}$ with the only nontrivial $2\times 2$ block $M'$ appearing at positions $2i-1$ and $2i$. Similarly a two-mode Gaussian operation acting on modes $i$ and $j$ will only have nontivial elements appearing at positions $2i-1$, $2i$, $2j-1$ and $2j$.

Below we list the single- and two-mode Gaussian operations. We first define each operation by its unitary representation $U$ with its action on the state given by $\ket{\Psi}\mapsto U \ket{\Psi}$  (the Schr\"odinger picture). Then we compute its action $o \mapsto U^\dag o \, U$  on the ladder and quadrature operator (Heisenberg picture). Finally we obtain the representation of the operation it terms of the pair $\bm d$ and $M$. We only specify the nontrivial block of $M$ and the nonzero elements of $\bm d$. 
\medskip

\paragraph{Phase shifter --} The phase shifter ($\Phi$) is a single mode operator given by the unitary operator $U_{\Phi}(\theta) = \exp(-i\theta  a^\dag  a)$.
When applied on the bosonic operators it gives
\begin{align}
    a &\mapsto e^{-i\theta} a \qquad \hspace{86 pt} a^\dag\mapsto  e^{i\theta}a^\dag \\
    \hat  x &\mapsto \cos(\theta)  \hat  x + \sin(\theta) \hat  p \hspace{53 pt}  \hat p \mapsto \cos(\theta)  \hat  p - \sin(\theta)  \hat  x.
\end{align}
We thus find that the corresponding symplectic transformation is characterized by
\begin{equation}
   \bm d=\vec{0} \qquad\qquad  M= \left(\begin{array}{cc}
    \cos(\theta) & \sin(\theta)\\
    -\sin(\theta) & \cos(\theta)
    \end{array}\right).
\end{equation}
\medskip

\paragraph{Displacement --} The unitary operator associated to a displacement (D) with amplitude $\alpha\in\mathbb{C}$ is given by $
    U_\text{D}(\alpha) = \exp(\alpha  a^\dag - \alpha^*  a)$.
This operation shifts the ladder and quadrature operators according to
\begin{align}
        a &\mapsto a + \alpha \hspace{87 pt} a^\dag \mapsto a^\dag + \alpha^* \\
        \hat x &\mapsto  \hat x+\Re\{\alpha\} \hspace{70 pt}  \hat p \mapsto  \hat p+\Im\{\alpha\}.
\end{align}
The symplectic transformation of a displacement  is hence characterized by
\begin{equation}
   \bm d = \left(\begin{array}{c}\Re\{\alpha\} \\ \Im\{\alpha\}\end{array}\right) \qquad \qquad M=\mathds{1}_2.
\end{equation}
We denote $\text{D}^x,\text{D}^p$ the displacement operations in the direction $\hat x,\hat p$ respectively. A displacement in $\hat x$ can be seen as a displacement with $\alpha$ real. Similarly, a displacement in $\hat p$ is a displacement with $\alpha$ imaginary.
\medskip

\paragraph{Single-mode squeezer --} The single-mode squeezing operation (S) is given by $U_\text{S}(z)=\exp\left(\frac{1}{2}\left(z^* a^2 -z \left(a^\dag\right)^2\right)\right)$,
where the complex parameter $z$ can be written as $z=r e^{i\theta}$. Single-mode squeezing transforms the ladder and quadrature operators as
\begin{align}
        a &\mapsto \cosh(r)  a - e^{i\theta}\sinh(r)  a^\dag  \hspace{115 pt}
         a^\dag \mapsto \cosh(r)  a^\dag - e^{-i\theta}\sinh(r) a \\
         \hat x &\mapsto \left(\cosh(r)-\sinh(r)\cos(\theta)\right)  \hat x - \sinh(r)\sin(\theta)  \hat p \hspace{30 pt}
         \hat p \mapsto \left(\cosh(r)+\sinh(r)\cos(\theta)\right)  \hat p - \sinh(r)\sin(\theta)  \hat x.
\end{align}
Hence, single-mode squeezing can be expressed by the symplectic transformation characterized by 
\begin{equation}
\bm d=\vec{0} \qquad \quad M = \cosh(r)\mathds{1}_2 - \sinh(r)\left(\begin{array}{cc}
    \cos(\theta) & \sin(\theta) \\
    \sin(\theta) & -\cos(\theta)
    \end{array}\right).
\end{equation}
\medskip

\paragraph{Two-mode squeezer --} A two-mode squeezer (TMS) acting on modes $i,j$ is defined by the unitary 
$U_\text{TMS}(z)=\exp(z^*  a_i  a_j -z  a^\dag_i  a^\dag_j),$
with $z=r e^{i\theta}$. This changes the bosonic operators as follows
\begin{align}
\begin{split}
         a_i &\mapsto \cosh(r)  a_i - e^{i\theta} \sinh(r)  a_j^\dag \hspace{74 pt}
        a_i^\dag \mapsto \cosh(r)  a_i^\dag - e^{-i\theta} \sinh(r) a_j \\
         a_j &\mapsto \cosh(r)  a_j - e^{i\theta} \sinh(r)  a_i^\dag \hspace{74 pt}
         a_j^\dag \mapsto \cosh(r)  a_j^\dag - e^{-i\theta} \sinh(r)  a_i.    
\end{split}
\\
   \begin{split}
         \hat x_i &\mapsto \cosh(r) \hat x_i - \sinh(r)(\cos(\theta) \hat x_j+\sin(\theta)  \hat p_j) \hspace{10 pt}
         \hat p_i \mapsto \cosh(r) \hat p_i + \sinh(r)(\cos(\theta)  \hat p_j-\sin(\theta)  \hat x_j) \\
         \hat x_j &\mapsto \cosh(r) \hat x_j - \sinh(r)(\cos(\theta)  \hat x_i+\sin(\theta)  \hat p_i) \hspace{10 pt}
         \hat p_j \mapsto \cosh(r) \hat p_j + \sinh(r)(\cos(\theta) \hat p_i-\sin(\theta)  \hat x_i).
    \end{split}      
\end{align}
The corresponding symplectic transformation is given by
\begin{equation}
    \bm d=\vec{0} \quad \quad M = 
    \left(\begin{array}{cccc}
    \cosh(r) & 0& -\sinh(r)\cos(\theta) & -\sinh(r)\sin(\theta)\\
    0& \cosh(r) & -\sinh(r)\sin(\theta) & \sinh(r) \cos(\theta)\\
    -\sinh(r)\cos(\theta) & -\sinh(r)\sin(\theta) &  \cosh(r) & 0 \\
     -\sinh(r)\sin(\theta) & \sinh(r) \cos(\theta) & 0& \cosh(r)
    \end{array}\right)    
\end{equation}
\medskip

\paragraph{Beamsplitter --} A beamsplitter (BS) on modes $i,j$ is given by the unitary $U_\text{BS}(\theta) = \exp(\theta( a^\dag_i  a_j -  a_i  a^\dag_j)),$
where the transmitivity is given by $\cos^2(\theta)$ and the reflectivity is $\sin^2(\theta)$. 
This Gaussian operation maps the operators to 
\begin{align}
    \begin{split}
         a_i &\mapsto \cos(\theta)  a_i + \sin(\theta)  a_j \hspace{80 pt}
         a_i^\dag \mapsto \cos(\theta) a_i^\dag + \sin(\theta) a_j^\dag 
        \\
         a_j &\mapsto \cos(\theta) a_j - \sin(\theta)  a_i \hspace{80 pt}
         a_j^\dag \mapsto \cos(\theta)  a_j^\dag - \sin(\theta)  a_i^\dag
    \end{split}
    \\
     \begin{split}
       \hat  x_i &\mapsto \cos(\theta) \hat x_i + \sin(\theta)\hat x_j \hspace{80 pt}
        \hat p_i \mapsto \cos(\theta) \hat p_i + \sin(\theta) \hat p_j \\
        \hat x_j &\mapsto \cos(\theta)\hat x_j - \sin(\theta)\hat x_i \hspace{80 pt}
       \hat  p_j \mapsto \cos(\theta)\hat p_j - \sin(\theta)\hat p_i.
    \end{split}
\end{align}
The corresponding symplectic transformation is characterized by
\begin{equation}
\bm d =\vec{0}\quad \quad
     \quad  M = \left(\begin{array}{cccc}
    \cos(\theta) & 0 & \sin(\theta) & 0 \\
    0 & \cos(\theta) & 0 & \sin(\theta) \\
    -\sin(\theta) & 0 & \cos(\theta) & 0 \\
    0 & -\sin(\theta) & 0 & \cos(\theta) 
    \end{array}\right).
\end{equation}

We note that the transformations D, S and TMS with arbitrary complex parameters $\alpha$ and $z=r e^{\ii \theta}$ can be decomposed as the same transformations with real parameters ($\alpha=\alpha^*$ and $z=r$) combined with two phase shifter.

\subsection{Measuring a Gaussian state with single photon detectors}
We here consider the measurement of one or several modes in a multimode state with non-photon number resolving (NPNR) detectors. The positive operator-valued measured (POVM) associated to the event "no-click" of a NPNR detection with efficiency $\eta$ operating on a mode with bosonic operators $a$ and $a^\dag$ is given by  $R^{a^\dag a}$ where $R=(1-\eta)$ while the event "click" is obviously related to the POVM $\id-R^{a^\dag a}$. Although such a measurement is not a Gaussian operation, we show that the multimode state conditioned on the outcome of such a measurement on one or several modes can be written as a mixture of Gaussian states and hence, its Wigner function can be written as a difference between two densities of a normal multivariate normal distribution. 
\medskip

\paragraph{Tracing out a mode ---} Let $W_\rho({\bm \alpha})$ the Wigner function of a $n$-mode state $\rho$. When the mode $i$ is traced out, the resulting state $\rho_{\lnot 1}$ is given by 
\begin{equation}
    \label{eq: ptrace}
    \begin{split}
    \rho_{\lnot i}&= \tr_i \rho\\
    &= \tr_i \, \pi^n \int d^2{\bm \alpha} W({\bm \alpha})\,  \bigotimes_{j=1}^n \delta( a_j-\alpha_j)\\
    &= \, \pi^{n-1} \int d^2{\bm \alpha} W({\bm \alpha})\,  \left(\tr_i  \pi \delta( a_i-\alpha_i)\right)\bigotimes_{\substack{j=1 \\ j\neq i}}^n \delta(a_j-\alpha_j)\\
    & = \pi^{n-1} \int d^2{\bm \alpha} W({\bm \alpha})\,\bigotimes_{\substack{j=1 \\ j\neq i}}^n \delta( a_j-\alpha_j).
\end{split}
\end{equation}
The second equality is obtained from the definition given in Eq.~\eqref{eq: Wigner multi} while the third inequality uses $\tr \delta( a-\alpha) = \pi$, see \cite{Vogel2006}. This shows that when the mode $i$ is traced out, the Wigner function of the remaining modes is simply given by  
\begin{equation}
W_{\lnot i}(\overline{\bm \alpha}) = \int d^2\alpha_i W({\bm \alpha}),
\end{equation}
that is, $\overline{\bm \alpha}$ is obtained from the vector $\bm \alpha$ by removing the components $2 i$ and $2i+1$. Since the marginals of a multivariate normal distribution are also normal, Gaussianity is preserved by the trace, that is $\rho_{\lnot i}$ remains Gaussian if $\rho$ is Gaussian.

Concretely, the displacement vector $\mu_{\lnot i}$ of  $\rho_{\lnot i}$ can be obtained by removing the components $2i$ and $2i+1$ of the displacement vector of $\rho$. Its covariance matrix $\Sigma_{\lnot i}$ is obtained by removing the rows and columns $2i$ and $2i+1$ of the covariance matrix of $\rho$.
\medskip

\paragraph{Outcome probabilities---} Let us consider a NPNR detector operating on a single mode with bosonic operators $ a$ and $ a^\dag$ and state $\rho$. The  probability of having a "no click" is given by
\begin{equation}
    \label{eq: PD}
    \begin{split}
        p_\text{no-click} &= \text{tr}\, \rho \, R^{ a^\dag  a} \\
        &= \int d^2\alpha W_\rho(\alpha) \underbrace{\tr \pi \delta( a-\alpha) R^{a^\dag a}}_{f_R(\alpha)}.
    \end{split}
\end{equation}
As $\delta( a- \alpha)$ can be written as $\delta( a- \alpha)=  \frac{1}{\pi^2}  \int d^2 \beta \,  e^{\alpha \beta^* - \alpha^*\beta } \t{D}(\beta)$~\cite{Vogel2006} with $\t{D}(\beta)$ the displacement operator with amplitude $\beta$, we have
\begin{equation}
    \label{eq: fra}
    \begin{split}
        f_R(\alpha)& =\frac{1}{\pi} \int d^2\beta \, e^{\alpha \beta^*-\alpha^* \beta}\tr R^{ a^\dag  a} D(\beta) \\
        &=\frac{1}{\pi} \int d^2\beta \, e^{\alpha \beta^*-\alpha^* \beta} e^{|\beta|^2/2} \tr  R^{ a^\dag  a} e^{-\beta^*  a} e^{\beta  a^\dag}   \\
        &=\frac{1}{\pi} \int d^2\beta \, e^{\alpha \beta^*-\alpha^* \beta} e^{|\beta|^2/2} \tr   e^{\beta  a^\dag} R^{a^\dag a} e^{-\beta^*  a}  \\
        &=\frac{1}{\pi^2} \int d^2\beta  \, e^{\alpha \beta^*-\alpha^* \beta}  e^{|\beta|^2/2}  \int d^2\gamma \, e^{\beta \gamma^*} e^{(R-1) |\gamma|^2} e^{-\beta^* \gamma} \\
        &=\frac{1}{\pi(1-R)} \int d^2\beta e^{\alpha \beta^*-\alpha^* \beta} e^{-\frac{1}{2}\frac{1+R}{1-R} |\beta|^2 } \\
        &=\frac{2}{(1+R)} e^{-2 |\alpha|^2 \frac{1-R}{1+R}}
    \end{split}
\end{equation}
where we used a writing of $D(\beta)$ as $e^{|\beta|^2/2} e^{-\beta^*  a} e^{\beta  a^\dag}$ is the second equality, the cyclic property of the trace in the third equality and the writing of $R^{ a^\dag  a}$ in the normal order $:e^{(R-1) a^\dag  a}:$ in the fourth equality. We deduce that the probability for having a "no-click" for any monomode state $\rho$ can be computed from its Wigner function as
\begin{equation}
    P_\text{no-click} = \int d^2\alpha W_\rho(\alpha) f_R(\alpha) \quad \text{with} \quad f_R(\alpha)=\frac{2}{(1+R)} e^{-2 |\alpha|^2 \frac{1-R}{1+R}}.
\end{equation}
\medskip

\paragraph{Heralding on photon detections ---} We now consider the case with $n$ modes with a photon detection on mode $i$. The sub-normalized state resulting of a "no-click event" on mode $i$ is given by
\begin{equation}
    \begin{split}
        \tilde \rho_{\circ i} &= \tr_i R^{ a_i^\dag  a_i} \rho\\
        &= \tr_i  R^{ a_i^\dag  a_i} \, \pi^n \int d^2{\bm \alpha} W({\bm \alpha})\bigotimes_{j=1}^n \delta( a_j-\alpha_j)\\
        & = \pi^{n-1} \int d^2{\bm \alpha} W({\bm \alpha})\left(\tr_i \pi R^{a_i^\dag a_i} \delta( a_i-\alpha_i)\right)\bigotimes_{\substack{j=1 \\ j\neq i}}^n \delta(a_j-\alpha_j)\\
        & = \pi^{n-1} \int d^2{\bm \alpha} W({\bm \alpha})f_R(\alpha_i) \bigotimes_{\substack{j=1 \\ j\neq i}}^n \delta( a_j-\alpha_j),
\end{split}
\end{equation}
where $f_R$ is defined in Eq.~\ref{eq: fra}. From Eq.~\ref{eq: Wigner multi}, the corresponding sub-normalized Wigner function is
\begin{equation}\label{eq: sub-normalized}
    \tilde W_{\circ i}(\overline{\bm \alpha}) = \int d^2\alpha_i W({\bm \alpha}) f_R(\alpha_i),
\end{equation}
with $\overline{\bm \alpha}$ the vector with $(n-1)$ elements constructed by dropping the $i$-th element of the vector ${\bm \alpha}$. Let us compute the normalisation for Gaussian states. We have
\begin{equation}\label{eq: norm1}
    \begin{split}
        \tilde W_{\circ i}(\overline{\bm \alpha})&= \int d\tilde{\balpha_{\bm i}} \frac{\exp\Big(-\frac{1}{2} (\tilde{\balpha}- \bm \mu)^T  \Sigma^{-1} (\tilde{\balpha}- \bm \mu) \Big)}{\sqrt{\text{det}(2\pi\Sigma)}} f_R(\tilde{\balpha_{\bm i}}),
    \end{split}
\end{equation}
where we denoted $\tilde{\balpha_i}=\begin{pmatrix}\Re{\alpha_i} \\ \Im{\alpha_i}\end{pmatrix}$. From Eq.~\ref{eq: fra}, we have 
\begin{equation}
    \begin{split}
        f_R(\tilde{\balpha}_{\bm i}) &= \frac{2}{ (1+R)} \exp(- |\tilde{\balpha}_{\bm i}|^2 \frac{2(1-R)}{1+R}) \\
        &= \frac{2}{(1+R)}  \exp\left(-\frac{2(1-R)}{1+R}\tilde{\balpha}_{\bm i}^T \tilde{\balpha}_{\bm i}\right)\\
        &= \frac{2}{(1+R)} \exp\left(-\frac{1}{2}\tilde{\balpha}^T F \tilde{\balpha}\right)
    \end{split}
\end{equation}
where $F$ is a matrix composed of $n-1$ blocks $F_j$, each block being a $2 \times 2$ block, such that
\begin{equation}
    F_{j\neq i} = \begin{pmatrix}0 & 0 \\ 0 & 0\end{pmatrix}, \qquad F_{j=i} = \frac{4(1-R)}{1+R}\mathds{1}_2.
\end{equation}
Eq.~\ref{eq: norm1} can therefore be written as
\begin{equation}
    \label{eq: Wnc1 int}
        \tilde W_{\circ i}(\overline{\bm \alpha}) = \frac{2}{ (1+R)\sqrt{\det (2\pi \Sigma)}}  \int d {\btalpha}_{\bm i} \exp\Big(-\frac{1}{2} (\tilde{\balpha}- {\bm \mu})^T  \Sigma^{-1} (\tilde{\balpha}- {\bm \mu}) - \frac{1}{2} \tilde{\balpha}^T F \tilde{\balpha} \Big) 
\end{equation}
 The integral can be rewritten as an integral of a multivariate distribution with a constant factor. To do so, we start by expressing the term in the exponent as
\begin{equation}
    \begin{split}
    &\frac{1}{2} (\btalpha - {\bm \mu})^T  \Sigma^{-1} (\btalpha - {\bm \mu}) + \frac{1}{2} \btalpha^T F \btalpha \\
    &= \frac{1}{2} (\btalpha- {\bm w})^T  (\Sigma^{-1}+F) (\btalpha - {\bm w}) + \frac{1}{2}{\bm \mu }^T \Sigma^{-1} {\bm \mu } -\frac{1}{2} {\bm w }^T (\Sigma^{-1}+F) \bm w
\end{split}
\end{equation}
with ${\bm w} = (\Sigma^{-1}+F)^{-1}\Sigma^{-1}\,{\bm \mu}$. This leads to
\begin{equation}
    \begin{split}
        \exp&\Big(-\frac{1}{2} ({\btalpha- \bm \mu})^T  \Sigma^{-1} ({\btalpha- \bm \mu}) - \frac{1}{2} {\btalpha}^T F {\btalpha} \Big)  \\
        & =\exp\Big(-\frac{1}{2} ({\btalpha- \bm w})^T  (\Sigma^{-1}+F) ({\btalpha- \bm w}) \Big) \exp\Big(-\frac{1}{2}{\bm \mu }^T \Sigma^{-1} {\bm \mu } +\frac{1}{2} {\bm w }^T (\Sigma^{-1}+F) \bm w\Big) \\ 
        &=N(\balpha ; {\bm w}, (\Sigma^{-1}+F)^{-1})\, \frac{\exp\Big(-\frac{1}{2}{\bm \mu }^T \Sigma^{-1} {\bm \mu } +\frac{1}{2} {\bm w }^T (\Sigma^{-1}+F) \bm w\Big)}{\sqrt{\det\big((\Sigma^{-1}+F)/2\pi\big)}},
    \end{split}
\end{equation}
where we used the identity $\det(2\pi X) = 1/\det(X^{-1}/2\pi)$.
Eq.~\ref{eq: Wnc1 int} can now be rewritten as
\begin{equation}
    \tilde W_{\circ i}(\overline{\bm \alpha}) = \frac{2}{1+R}\sqrt{\frac{1}{\det(\Sigma) \det(\Sigma^{-1}+F)}} 
    e^{-\frac{1}{2}{\bm \mu }^T \Sigma^{-1} {\bm \mu } +\frac{1}{2} {\bm w }^T (\Sigma^{-1}+F) \bm w}
    \int d^2{\alpha}_{i}\, N(\bm \alpha ; {\bm w}, (\Sigma^{-1}+F)^{-1}).
\end{equation}
Note that the marginal of the multi-variate normal distribution is normalized. We deduce that the properly normalized Wigner function $W_{\circ i}(\balpha)$, which can be written as $\tilde W_{\circ i}(\balpha) = p_{\circ i} W_{\circ i}(\balpha)$
is given by 
\begin{equation}
    W_{\circ i}(\overline{\bm \alpha})  = N(\overline{\bm \alpha}; \bm \mu', \Sigma')
\end{equation}
where the displacement vector ${\bm \mu}'$ is obtained by removing the elements $2i$ and $2i+1$ of $(\Sigma^{-1}+F)^{-1} \Sigma^{-1} {\bm \mu}$ and the covariance matrix $\Sigma'$ is obtained by removing the raws and column $2i$ and $2i+1$ of $(\Sigma^{-1} + F)^{-1}$. $p_{\circ i}$, the probability of a no-click outcome when applying a NPNR detector on mode $i$, is given by 
\begin{equation}
    p_{\circ i}  = \frac{2}{1+R}\sqrt{\frac{1}{\det(\Sigma) \det(\Sigma^{-1}+F)}}  
    e^{-\frac{1}{2}{\bm \mu }^T\left( \Sigma^{-1} - \Sigma^{-1} (\Sigma^{-1}+F)^{-1} \Sigma^{-1} \right){\bm \mu }}.
\end{equation}

We can finally express the state $\rho_{\bullet i}$ conditioned on a "click" on mode $i$ by considering its connection with the subnormalized state
\begin{equation}
     \tilde \rho_{\bullet i} = \tr_i \rho (\mathds{1}-R^{a_i^\dag a_i}) =\rho_{\lnot i} - p_{\circ i} \rho_{\circ i} =p_{\bullet i} \rho_{\bullet i}
\end{equation}
where $p_{\bullet i}=1-p{\circ i}$ is the probability of having a click on mode $i$. We deduce
\begin{equation}
\rho_{\bullet i} = \frac{\rho_{\lnot i} - p_{\circ i} \rho_{\circ i}}{1 - p_{\circ i}}
\end{equation}
and the Wigner function of this normalized state is given by
\begin{equation}
    \label{eq: Wigner cond}
    W_{\bullet i}(\overline{\bm \alpha}) = \frac{W_{\lnot i}(\overline{\bm \alpha}) -p_{\circ i}W_{\circ i}(\overline{\bm \alpha}) }{{1 - p_{\circ i}}}=  \frac{W_{\lnot i}(\overline{\bm \alpha}) -\tilde W_{\circ i}(\overline{\bm \alpha}) }{{1 - p_{\circ i}}}.
\end{equation}
This shows that the Wigner function of the conditional state $\rho_{\bullet i}$ can be written as a weighted sum of Gaussian Wigner functions. 
Note that Gaussian operations acting on the conditional state can be accounted by first considering their actions on individual Gaussian Wigner functions and by then recombining the two branches according to the weight sum of initial Wigner functions.
\medskip

\paragraph{Statistics of NPNR detections on multiple modes ---}
We finally consider the detection of $m$ modes with NPNR detectors and show the way to compute the probability of outcomes. We represent an arbitrary outcome by a vector $\bf k$ where the i$^th$ component $K_i$ equals $0$ for a no-click event and $1$ for a click. The probability of getting such an outcome is given by 
\begin{equation}
\mean{E_{\bf k}} = \mean{\bigotimes_{i|k_i=0} R^{a_i^\dag a_i} \bigotimes_{j|k_j=1}(\mathds{1} -R^{a_j^\dag a_j})}.
\end{equation}
According to Eq.~\eqref{eq: PD}, we have 
\begin{equation}
\label{statisticsEk}
\begin{split}
\mean{E_{\bf k}}  = \int d^2\bm \alpha \, W_\rho(\bm \alpha ) \sum_{\bm \ell \in S_{\bm k}} (-1)^{\bm k\cdot\bm \ell} \prod_{i=0}^n \left(f_R(\alpha_i)\right)^{l_i} 
 = \sum_{\bm \ell \in S_{\bm k}} (-1)^{\bm k\cdot\bm \ell}  \mean{\hat{o}_{\bm \ell}}
\end{split}
\end{equation}
where $S_{\bm k}$ is the set containing all strings $\bm \ell$ with $n$ bits where the components $\ell_i$ are fixed to $1$ for all $i$ such that $k_i=0$. This set contains $2^{|\bm k|}$ terms with $|\bm k|= \sum_i k_i$. From the value of $f_R(\alpha)$ given in Eq.~\eqref{eq: fra}, we have
\begin{equation}
\label{oelle}
    \begin{split}
    \mean{\hat{o_{\bm \ell}}} & = \left(\frac{2}{1+R}\right)^{|\bm \ell|}\int d^2\balpha \, W(\bm \alpha ) \exp\left(-\frac{2(1-R)}{1+R}\sum_i \ell_i |\alpha_i|^2 \right) \\
    & = \left(\frac{2}{1+R}\right)^{|\bm \ell|}\int d^2\balpha \, W(\bm \alpha ) \exp\left( \frac{1}{2} \btalpha^T O_{\bm \ell} \btalpha\right)\\
    &= \left(\frac{2}{1+R}\right)^{|\bm \ell|} \int d^2\balpha \, N(\bm \alpha;\bm \mu ,\Sigma) \, e^{-\frac{1}{2}\btalpha^T O_{\bm \ell} \btalpha}\\
        &= \left(\frac{2}{1+R}\right)^{|\bm \ell|} \frac{\exp\left(-\frac{1}{2}\bm \mu^T (\Sigma^{-1} - \Sigma^{-1}(\Sigma^{-1}+O_{\bm \ell})^{-1}\Sigma^{-1} )\bm \mu \right)}{\sqrt{\det(\Sigma)\det(\Sigma^{-1}+O_{\bm \ell})}}.
    \end{split}
\end{equation}
where $|\bm \ell|=\sum_i \ell_i$ and $O_{\bm \ell}$ the block diagonal matrix
\begin{equation}
O_{\bm \ell} = \frac{4(1-R)}{1+R} \bigoplus_{i=1}^m \ell_i \begin{pmatrix}
1 & 0\\
0& 1
\end{pmatrix}.
\end{equation}
The last equality is obtained by noting that 
\begin{equation}
(\btalpha -\bm \mu)^T \Sigma^{-1} (\btalpha -\bm \mu) + \btalpha^T O_{\bm \ell} \btalpha =  ({\btalpha- \bm w})^T  (\Sigma^{-1}+O_\ell) ({\btalpha- \bm w}) + {\bm \mu }^T \Sigma^{-1} {\bm \mu } - {\bm w }^T (\Sigma^{-1}+O_\ell) {\bm w}
\end{equation}
for ${\bm w} =(\Sigma^{-1}+O_\ell)^{-1}\Sigma^{-1} \,{\bm \mu}$ and 
\begin{equation}
\int d^2\balpha N(\bm \alpha;\bm w ,(\Sigma^{-1}+O_\ell)^{-1})=1.
\end{equation}
The expected statistics is obtained by combining~\eqref{statisticsEk} and~\eqref{oelle}.

\medskip

\section{Appendix B. Automated design implementation}

We here provide a description of how we automatise the design of quantum photonics experiments for DIQKD\@. This automatization is based on reinforcement learning, a sub-field on machine learning.
Reinforcement learning (RL) algorithms aim at finding out what action an \textit{agent} should take when interacting with its \textit{environment} in order to maximize the cumulative reward, see Fig.~\ref{fig:RL_Overview}.
Here, we give a quick overview of policy gradients and the proximal policy optimisation algorithm -- the type of agent we used. Then we dive in the details of our implementation before concluding with its convergence.

\begin{figure}[ht]
    \centering
    \includegraphics[width=.90\textwidth]{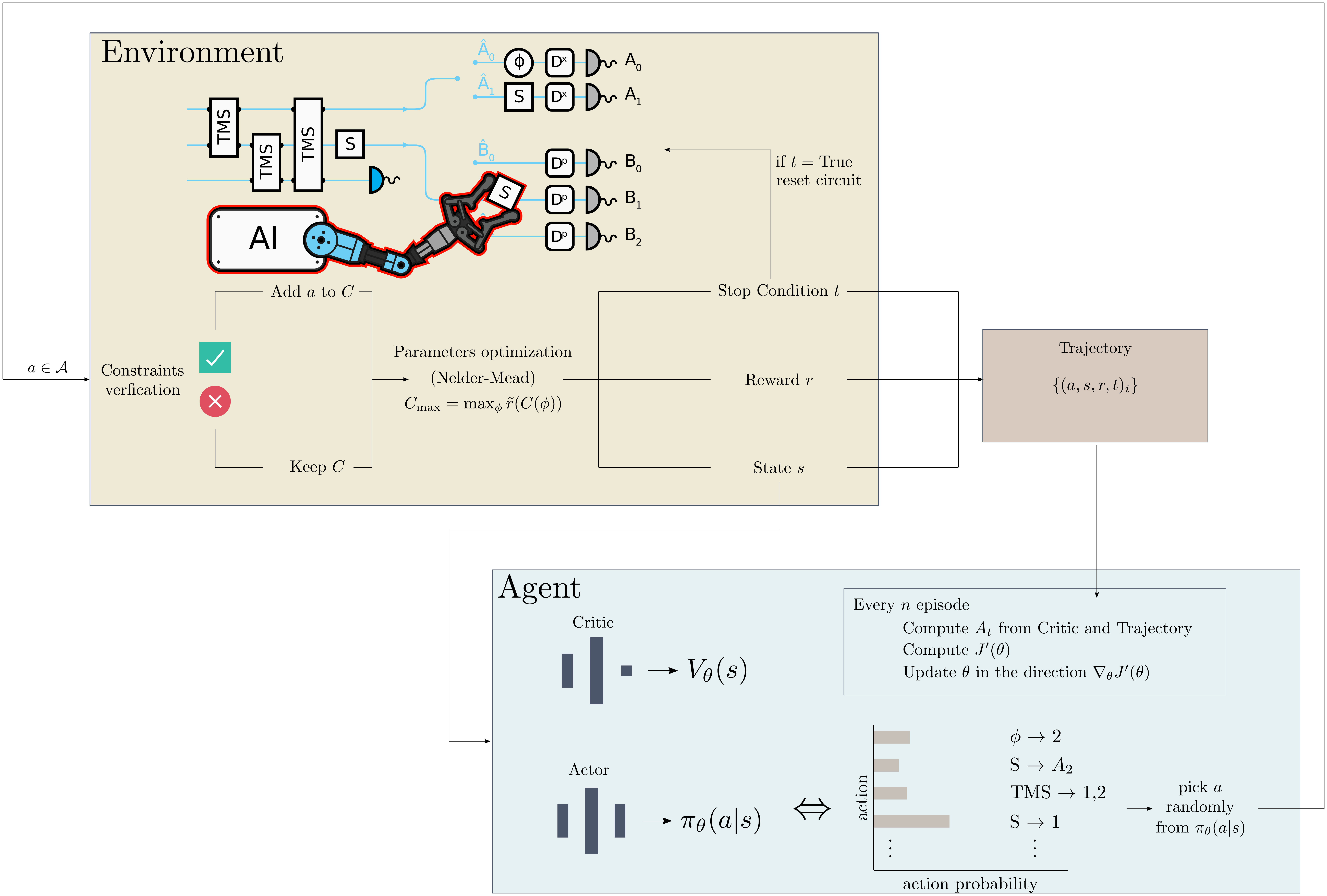}
    \caption{Reinforcement learning for automated circuit design. An action $a$ is send to the environment. If it verifies some constraints (avoid redundancy, avoid beam splitter on the vacuum, etc..) it is added to the optical circuit $C$, otherwise $C$ is unchanged. The circuit's parameters are then optimised. A resulting reward $r$, state $s$ and stop condition $t$ is passed both to the agent and to a memory or trajectory. If the stop condition is \textit{true} the circuit is reset to empty modes and the episode ends. Otherwise, for $s$ and the policy $\pi_\theta$ the agent picks a new action and sends it to the environment. Every $n$ episode, the agent uses the trajectory to update the parameters $\theta$ to maximize future rewards.}\label{fig:RL_Overview}
\end{figure}

\medskip

We emphasize that our algorithm is inspired by the one presented in Ref. \cite{Melnikov20}. There, it has shown how to automate the search of optical experiments for winning non-local games using projective simulation, simulated annihilating and a numerical framework based on Fock representation. Our approach differs in all these aspects -- we used policy gradient based reinforcement learning, Nelder-Mead optimisation and a faster and more reliable custom-made numerical framework based on Gaussian representation of states to simulate optical circuits.

\smallskip

\subsection{Policy gradient overview}

 The behaviour of an agent during its interaction with the environment is captured by its (stochastic) \textit{policy} $\pi_\theta(a|s)$--  the probability distribution of taking action $a\in \mathcal{A}$ when perceiving a state $s$. Policy gradient methods are a type of RL algorithms that aim at learning the parameters $\theta$ of the policy $\pi_\theta(a|s)$  in order to find the strategy maximizing an estimate of the sum of total future rewards.
This learning occurs by sampling trajectories -- a sequence of state $s$, action $a$ and reward $r$ -- for a given policy, followed by an update of the policy's parameters that favors the trajectories with the highest cumulative reward.

For a given reward function $J_\theta(r,s)$, gradient ascent is used to update the parameters $\theta$ by a certain amount given by the learning step.
Note that this learning step is an hyperparamater crucial to training stability, e.g. a too small step  results in a stagnation of the policy whereas a too large step changes may hinders the convergence of the policy.


 For a concise overview of different policy gradient methods, consult Ref.~\cite{Weng2018}. Here, we note  that 
policy gradient algorithms can   also be used in model-free  RL. Model-free algorithms optimize the  policy without any knowledge   of the internal functioning of the environment, i.e.\@ the reward function and the transition function (the probability to obtain an output state given an input state and an action). This approach   is very versatile and  well-suited for environments with a complex transition model, e.g. if the state-transition depends on an internal complex stochastic process such as a nonlinear optimization as in our case, see below.

\smallskip

\subsection{Proximal policy optimisation}

 For our implementation we use a model-free policy gradient method known as proximal policy optimisation (PPO).  This algorithm is known to be sample efficient and has seen numerous successful applications while remaining fairly simple to implement and use.


PPO is an \textit{on-policy} algorithm, i.e.\@ policy updates are computed based on 
the latest policy $\pi_{\text{old}}$ (parameterized by some parameter value $\tilde \theta$) and a batch of trajectories $\{(\bm a_t, \bm s_t, \bm r_t)\}$ sampled from this policy. 
The \textit{clipped surrogate} reward function used by PPO reads~\cite{Schulman2017}
\begin{equation}
    \label{eq:PPO_rf}
    J^{\text{CLIP}}_\theta = \mathds{E}_t[\min(R_t(\theta)A_t(\theta),\, \text{clip}(R_t(\theta),1-\epsilon,1+\epsilon)A_t(\theta))].
\end{equation}
 Here, $R_t(\theta)=\pi_\theta({\bm a}_t|{\bm s}_t)/\pi_{\text{old}}({\bm a}_t|{\bm s}_t)$ is the probability ratio between a policy $\pi_\theta$ and the old policy $\pi_\text{old}$, $A_t(\theta)$ is the advantage function, $\epsilon$ is an hyperparameter and the expected value $\mathds{E}_t$ is taken over the batch of trajectories. The \textit{clip} function is used to constrain the amplitude of policy changes, quantified by $R_t(\theta)$, to be within the interval $[1-\epsilon,1+\epsilon]$. Specific to PPO, this limitation increases the robustness to sub-optimal learning step choices by preventing large policy updates.  The advantage function $A_t(\theta)$ quantifies how good was the choice of the action sequence $\bm a_t$ in reaction to the perceived states $\bm s_t$ when compared to a baseline. The baseline reward expected from a state $s_k$ is estimated with the \textit{value} function $V_\theta(s_k)$, see below. For each step $k$ of a given trajectory the advantage function thus compares the returned reward $r_k$ plus the reward predicted for the next state $V_\theta(s_{k+1})$ (resulting from the action $a_k$) to the predicted total baseline reward $V_\theta(s_k)$ for the observed state $s_k$. Formally, for a trajectory segment containing $T$ steps, the advantage is given by
\begin{equation}
       A_t =  \delta_t + (\gamma \lambda)\delta_{t+1} + \dots + (\gamma \lambda)^{T-t+1}\delta_{T-1} \qquad \text{with} \qquad \delta_k = r_k + \gamma V_\theta (s_{k+1})-V_\theta(s_k)
\end{equation}
 where $\lambda$ and $\gamma$ are hyperparameters that can be tuned to discount rewards that are delayed in time from the action, see \cite{Schulman2017} for details. In summary, the product $R_t(\theta)A_t(\theta)$ appearing in the reward function $J_\theta^{\text{CLIP}}$ is large if the choice of the action sequence $\bm a_t$ is advantageous when compared to the baseline ($A_t > 0$) and the policy $\pi_\theta$ is more likely to choose these actions than $\pi_\text{old}$ ($R_t>1$).

The value function $V_\theta(s)$  estimates the predicted cumulative reward obtained when following the policy $\pi_\theta$  starting from the state $s$. It shares the parameters $\theta$ with the policy, and is learned by minimizing the error-term
\begin{equation}
    \label{eq:PPO_vf}
    L^V_\theta = (V_\theta(s) - V^{\text{target}})^2,
\end{equation}
where $V^{\text{target}}$ is the computed value from the trajectory batch. Since $\theta$ is common to both the policy and the value function, this  error-term can simply be added to the reward function given in Eq.~\ref{eq:PPO_rf}~\cite{Schulman2017}.

Finally, to enhance the exploration of the search space, an entropic penalty term $H(\pi_\theta)$ is added to the reward function. It favors policies that are not deterministic, but explore different actions.

To summarize, the total PPO reward function used to optimize the policy for higher cumulative reward is
\begin{equation}
    \label{eq: loss}
    J_\theta =  J^{\text{CLIP}}_\theta + \mathds{E}_t[- c_1 L^V_\theta + c_2  H(\pi_\theta) ]
\end{equation}
where weights $c_1,c_2$ are hyperparameters. 

\smallskip

\paragraph{PPO implementation --}

We used an implementation of PPO available in the ReinforcementLearning.jl package~\cite{Tian2020}.
This implementation uses two neural networks -- an actor network and a critic network -- sharing parameters $\theta$.

The actor network acts as the policy. It takes as an input the state received from the environment. The output layer has one neuron for each of the possible actions.  The higher is the value of the activation function for one of these neurons the higher is the probability to choose the corresponding action.

The critic network acts as the value function $V_\theta$. It also takes the observed state as the input. The output layer is composed of a single neuron whose activation value is directly proportional to the outcome of $V_\theta$.


 We tested multiple hyperparameters configuration. We settled on a choice of a single hidden layer of 256 neurons for both neural networks. For the hyperparameters of the advantage function, we set the discount factor to $0.99$ and the smoothing parameter $\lambda$ to $0.95$. We consider learning from trajectory of size $T=32$. The reward function is parameterized by a clip range of $\epsilon=0.1$ and weights $c_1=0.5$ and $c_2=10^{-3}$.

\smallskip

\subsection{Interaction with the environment}

The PPO agent interacts with an environment that  plays the role of a virtual lab where the $n$-mode photonic setup is implemented and the DIQKD protocol is executed.
 As described in the main text, the photonic circuit   can be decomposed in two "phases". The state preparation phase, ending with the heralding measurements of all but the two first modes, which is followed by the measurement phase, where Alice \& Bob can each perform local operations on their mode before measuring it with an NPNR detector. Actions that are taken by the agent correspond to  placing optical elements on specific locations in the circuit.

A \textit{step} starts with the agent taking an action, then the environment updates the photonic circuit, optimise the circuit parameters, and return a corresponding reward, the updated state and a stop condition. This stop condition is a Boolean variable that becomes True when the total number of actions taken by the agent reaches a threshold we fixed at $15$. When the stop condition is True the circuit and the action counter are reset. An \textit{episode} is defined as the series of actions taken until the stop condition occurs.

\smallskip 

The optical elements we consider are displacements (D), phase shifters ($\Phi$), single-mode squeezers (S), two-mode squeezers (TMS) and beam-splitters (BS) operations. 
These are all Gaussian operations that are detailed in the previous section.
An action $a$ is composed of an optical element and a location on the circuit, i.e.\@ on which mode it acts on.
In addition, we denote $a(\phi)$ the action $a$ with its optical element paramaterized by $\phi$ which is either a real number, if $a$ is a phase shifter or a beam-splitter, or a complex number otherwise. 

In the state preparation phase, we allow for phase-shifters and squeezers to be applied on all of the $n$ modes. This results in the following set of actions
\begin{equation}
    \mathcal{A}_\text{prep} = \{\Phi^{(i)},\text{S}^{(i)},\text{TMS}^{(ij)},\text{BS}^{(ij)}\}
\end{equation}
with $i,j\in\{1,\dots,n\}$ specifying the modes on which they act. In the measurement phase, we consider Alice (Bob) to always perform a displacement along the $x$ ($p$) direction before the NPNR detector. In addition, Alice \& Bob can perform actions from the sets
\begin{equation}
    \begin{split}
        \mathcal{A}^\text{Alice}_\text{meas} &= \{\text{D}^{p}_0, \text{S}_0, \text{PS}_0,\text{D}^{p}_1, \text{S}_1, \text{PS}_1,\} \\
        \mathcal{A}^\text{Bob}_\text{meas} &= \{\text{D}^{x}_0, \text{S}_0, \text{PS}_0,\text{D}^{x}_1, \text{S}_1, \text{PS}_1,\text{D}^{x}_2, \text{S}_2, \text{PS}_2\},
    \end{split}
\end{equation}
Here, the subscripts denote the choice of the measurement setting following which the operation is performed (Alice \& Bob only receive a single mode).
Combining the three sets, we obtain the total set of all possible actions that can be taken in our environment
\begin{equation}
    \mathcal{A} = \mathcal{A}_\text{prep} \cup \mathcal{A}^\text{Alice}_\text{meas} \cup \mathcal{A}^\text{Bob}_\text{meas}.
\end{equation}

\subsection{Circuit Parameter Optimization}

A succession of $N$ actions $a_1,\dots, a_N$ from $\mathcal{A}$ with the corresponding parameter values $\phi_1,\dots, \phi_N$ defines a photonic circuit
\begin{equation}
    C(\bm \phi) = \{a_1(\phi_1),\ldots,a_n(\phi_N)\}, \quad \forall a_i \in \mathcal{A}. 
\end{equation}
In order to avoid trivial actions and redundancy, we add some constraints on such circuits. If a phase shifter or a beam splitter is added on an empty mode, i.e.\@ acting on the vacuum, we can simply discard the corresponding action from $C$.
If two identical actions $a$ are either consecutive or separated by actions that commute with $a$, the latter occurrence of $a$ is discarded from $C$.
Finally we constrained the actions in the measurement set $A_\text{meas}$ to be unique, i.e.\@ for each $a\in A_\text{meas}=\mathcal{A}^\text{Alice}_\text{meas} \cup \mathcal{A}^\text{Bob}_\text{meas}$ we discard any other occurrence of $a$ in $C$.

The circuit $C_{\max}$ is the circuit with parameters ${\bm \phi}_{\max}$ and noisy pre-processing probability optimized to maximize the key rate obtained from Eq.~\ref{eq:keyrate}. 
Such an optimization can be hard to performed since the key rate defined in Eq.~\ref{eq:keyrate} is only valid for CHSH score $S>2$.
To help the optimization to converge, we define the \textit{extended key rate} as the continuation of the key rate formula
\begin{equation}
    \label{eq: keyrate_extended}
    \tilde{r} = 1 - \tilde{I_p} - \entVN(\mathbf{A}'|\mathbf{B})\\
\end{equation}
with
\begin{equation}\small \nonumber
    \tilde{I_p} = \begin{cases}
        h \left( \frac{1+\sqrt{(|S|/2)^2-1}}{2} \right) - h\left(\frac{1+\sqrt{1-p(1-p)(8-|S|^2)}}{2}\right), & \text{if}\ |S|>2 \\
        1 + h\left(\frac{1+\sqrt{|S|/2}}{2}\right) - h\left(\frac{1+\sqrt{1-p(1-p)|S|^2}}{2}\right), & \text{otherwise}
        \end{cases}
\end{equation}
(see Fig.~\ref{fig:keyrate_extended}) to all possible values of the CHSH score, as plotted in Fig.~\ref{fig:keyrate_extended}. Note that $\tilde r$ is negative for local values of CHSH $|S|\leq2$. The extended key rate function is what we used to optimize the parameters of the circuit $C$, i.e.\@ to numerically solve  
\begin{equation}
    \bm \phi_{\max} = \text{argmax}_{\bm \phi}\,  \tilde{r}(C({\bm \phi}))
\end{equation}

Concretely, this optimisation was done using the Nelder-Mead method~\cite{Nelder1965}. For a new circuit, we use multiple random starting parameters.
However, for a circuit $C'$ constructed by adding a new action to a previously optimized circuit $C_{\max}$, we optimize the parameters of $C'$ starting from $({\bm \phi}_{\max} \, ,0)$. To avoid local optima, two additional optimizations with different starting points are performed, one from a random point and one from $({\bm \phi}_{\max}+\varepsilon {\bm u} \, ,0)$ where ${\bm u}$ is a uniformly distributed real vector with value in $[0,1]$ and $\varepsilon$ is a scalar we fixed to $0.2$.

In the case where $n\geq3$ some modes  are be used for heralding in the state preparation phase.
In this case, there exist circuits that  never produce the heralding event, e.g. when heralding on an empty (vacuum) mode.
Similarly, some circuit parameter values also render heralding impossible. 
In both of these case, we fix the key rate to a dummy value of $-1$.
This allows the minimization to run over the entire space of parameters and avoids the implementation of complicated parameter constraints.

We tested our circuit optimisation strategy on the reference photonic implementation of DIQKD shown Fig~\ref{fig:pola}. 
We were able to recover similar key rates than the ones derived analytically In Ref.~\cite{Ho2020}.
In particular, we found the same efficiency threshold for key rate higher than $10^{-9}$.
Below this order of magnitude, the optimization becomes too unstable and the numerical quantum optics simulation starts to induces non negligible error due to matrix inverse operations.

\begin{figure}[ht]
    \centering
    \includegraphics[width=.45\textwidth]{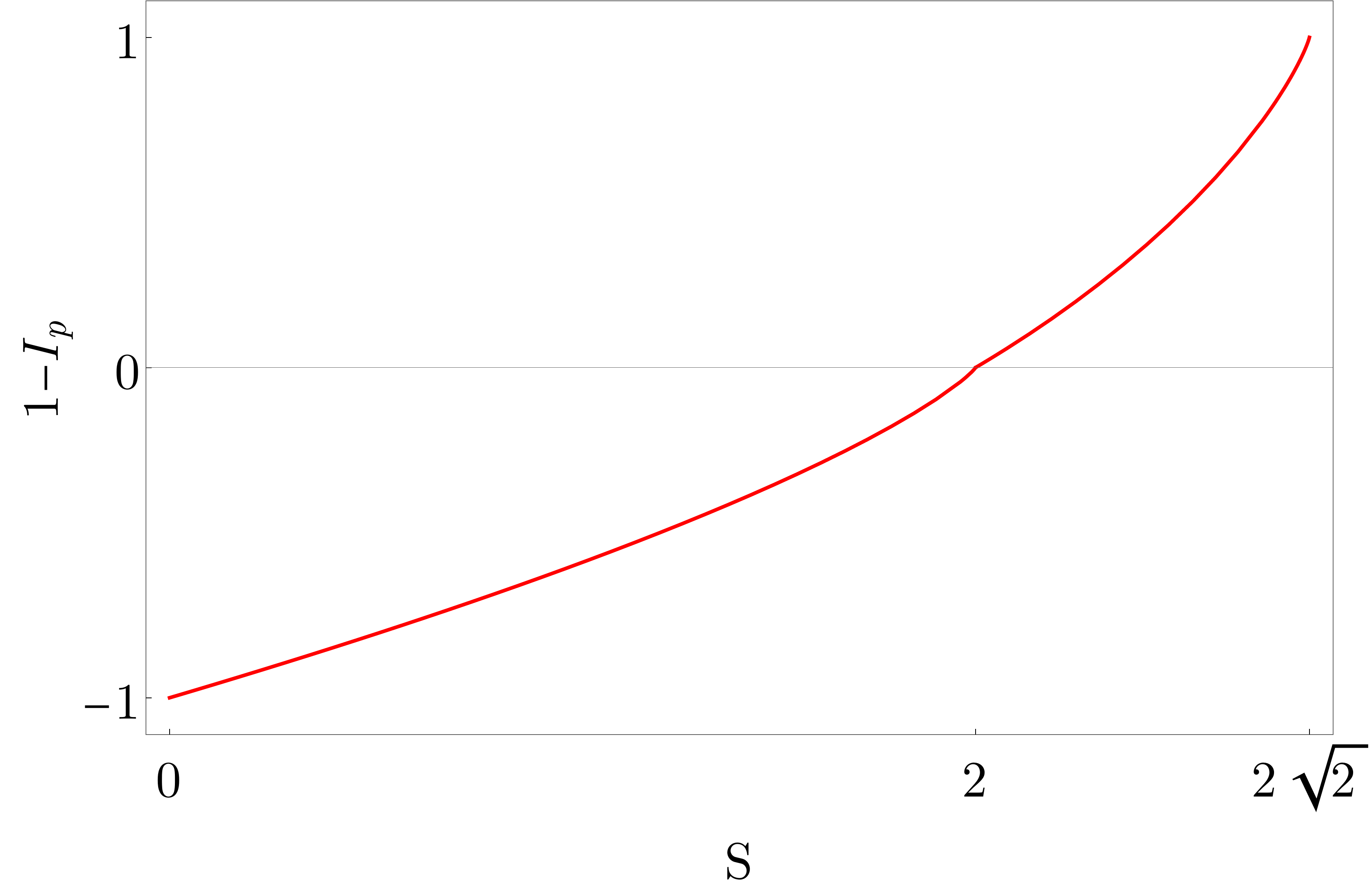}
    \caption{Evolution of the extended version of $1-I_p$ with the CHSH score $S$ in the absence of noisy pre-processing ($p=0$).}\label{fig:keyrate_extended}
\end{figure}

\subsection{Reward Function}

For a given circuit $C$, the environment computes a reward evaluating the performance of the circuit for a fixed task.
We investigated two tasks for i) finding circuits maximizing the key rate in the absence of loss, and ii) finding loss tolerant circuit, i.e.\@ maximizing loss while maintaining a key rate greater that $10^{-4}$.

In the first case, a naive approach would be to use the extended key rate $\tilde r$ in Eq.~~\ref{eq: keyrate_extended} as the reward
\begin{equation}
    r = \tilde{r}(C_{\max}).
\end{equation}
To help the PPO algorithm to converge, we reshape this reward using a secondary reward based on the CHSH value $S$ attained by $C_{\max}$
\begin{equation}
	\label{eq: reward 1}
    r = \frac{1}{1+\varepsilon}(\tilde{r}(C_{\max}) + \varepsilon\,  |S(C_{\max})|)
\end{equation}
where we fixed the weight $\varepsilon=10^{-2}$.

In the second case, the reward is the minimum efficiency so that the circuit can output a key rate greater than the threshold set to $10^{-4}$.
Denote $C^\eta$ the circuit with NPNR detectors of efficiency $\eta$.
To evaluate this reward, we start by optimising the circuit $C^\eta$ in the perfect case, i.e.\@ with efficiency $\eta=1$.
We then lower $\eta$ by a step $s$ depending on the order of magnitude of the key rate $\tilde{r}(C_{\max})$ following
\begin{equation}
    s = \max(2\times 10^{\lfloor{\log_{10}(\tilde{r}(C_{\max})-1}\lfloor},10^{-3}).
\end{equation}
For the newly obtained efficiency, we optimise the circuit again, starting from the optimal parameters found at the previous step. Eventually, $\eta$ is too small and the optimal key rate found by the optimization does not exceed the threshold. We label $\eta_{\min}$ the smallest efficiency  before this happens. The reward is then given by
\begin{equation}
	\label{eq: reward min}
    r = \eta_{\min}.
\end{equation}

Note that in both tasks, the PPO algorithm learns from the total reward gathered per episode. The reward attributed at step $i$ is thus defined as
\begin{equation}
    \begin{cases}
    r_i - r_{i-1}, & \text{if}\, r_i - r_{i-1} > 0 \\
    0, & \text{otherwise}
    \end{cases}
\end{equation}
so that only the highest reward obtained during an episode matters.

\subsection{The state perceived by the agent}

For each circuit $C$, the state $s$ returned by the environment to the PPO algorithm is taken from its optimised version $C_{\max}$ at the measurement phase, just before the NPNR detectors. 

For each setting pairs $\{x,y\}$ we extract the quantum state, i.e.\@ the covariance matrix $\Sigma$ and the displacement vector $\bm \mu$ for each Gaussian state composing the conditional state as in \ref{eq: Wigner cond}. Note that, thanks to symmetries of the Hermitian nature of the covariance matrix on $m$-modes, a single covariance matrix is parameterized by $(2m^2+m)$ real parameters. Denote $\vec{\Sigma}$ the vector containing these real parameters. For a Gaussian state, all parameters are contained in the vector
\begin{equation}
    \vec{p} = \begin{pmatrix}
    \vec{\Sigma} \\
    \bm \mu
    \end{pmatrix}.
\end{equation}
Trivially, an heralded state represented as a quasi-mixture of $2^{n-m}$ Gaussian states with weight $w_i$ can be represented by the vector
\begin{equation}
    \bm p = \begin{pmatrix}
    w_1 \\
    \vec{p}_1 \\
    \vdots \\
    w_{2^{n-m}} \\
    \vec{p}_{2^{n-m}}
    \end{pmatrix} = \begin{pmatrix}
    w_1 \\
    \vec{\Sigma}_1 \\
    \bm \mu_1 \\
    \vdots \\
    w_{2^{n-m}} \\
    \vec{\Sigma}_{2^{n-m}} \\
    \bm \mu_{2^{n-m}}
    \end{pmatrix}.
\end{equation}
Labelling $\bm p_{xy}$ the vector containing the information of the quantum state for the setting choice $x,y$, the state returned by the environment is the vector
\begin{equation}
    \bm P = \begin{pmatrix}
        \bm p_{00} \\
        \bm p_{01} \\
        \vdots \\
        \bm p_{12}
    \end{pmatrix}.
\end{equation}

\medskip

\subsection{PPO convergence}

To grasp the convergence of our algorithm, we investigated the evolution of different factors with learning steps.
The reward function, also called \textit{loss}, used to optimize the policy, Eq.~\ref{eq: loss}, should converge to zero, i.e.\@ showing a convergence of the learning algorithm and a successful descent of the gradient of this function.
When the policy converges to a more optimal one, the reward received from the environment should, on average, increase with learning steps.

The evolution of these quantities is depicted in Fig.~\ref{fig:RL_plots}. 
As an example, we choose a scenario where the PPO is aiming for quantum optical circuits maximizing key rates in a lossless scenario.
In this case, the reward is simply the key rate obtained for perfect detector efficiency. 
Furthermore, since the policy is stochastic, in order to get more relevant statistics, we trained the PPO on 10 environments simultaneously. That is, a single agent interacts with ten optical setups in parallel.
The reward in Fig.~\ref{fig:RL_plots} is the cumulative reward received from these environments.
Because of the stochastic nature of the learning process, it is relevant to study a smoothed evolution of the reward and loss with learning steps.
We choose to define the smoothed evolution $s$ of a quantity $x$ at step $t$ as
\begin{equation}
    \label{eq: smoothed}
    s(x,t) = \begin{cases}
    x(0), \quad \text{if}\, t=0 \\
    s(t-1)\omega +x(t)(1-\omega),\quad \text{otherwise}
    \end{cases} 
\end{equation}
for some weight $\omega\in[0,1]$.

In Fig.~\ref{fig:RL_plots}, we see three phases occurring during the policy optimisation.
First, an exploration phases, where the policy is close to random and where the reward obtained are relatively low. The loss is contained in a $[0,100]$ interval. 
Then, a drastic increase in the reward occur. This is first triggered by a high loss of around $250$. 
Finally, both the reward and the loss plateau, with loss getting close to $0$ and the average reward over the $10$ environments oscillating around $~0.84$. This can be interpreted as a convergence of the policy (low loss) in an relevant region (high reward).
Note that the spikes in the loss evolution are due to gradient descent on batches -- some batch of data will randomly contain episode with better reward than others.

\begin{figure*}[t]
    \centering
    \includegraphics[width=.95\textwidth]{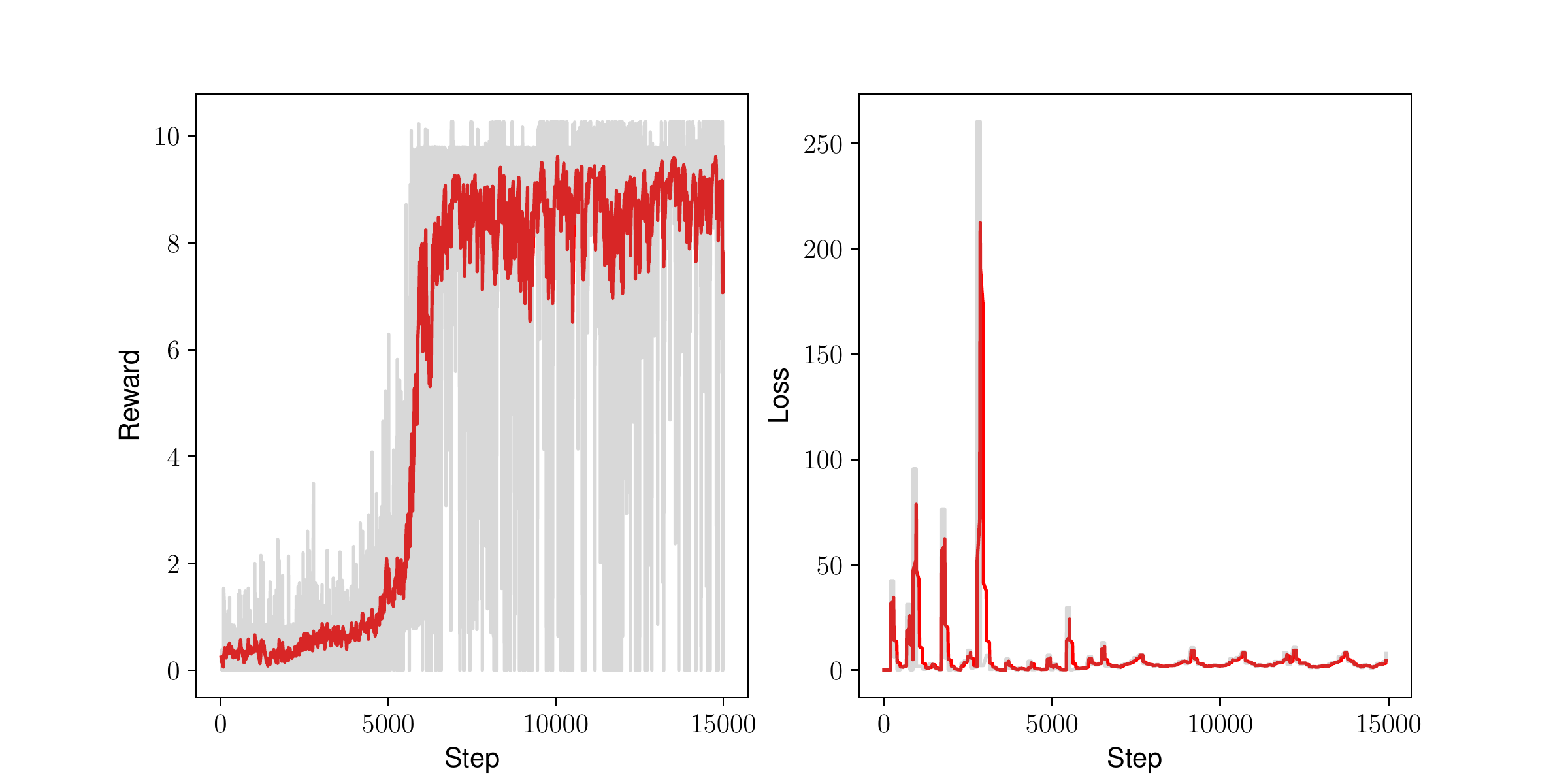}
    \caption{Evolution of the reward $r$ and loss $J_\theta$ with learning steps, in the case of PPO aiming to design quantum optical circuits maximizing the key rate in a lossless scenario.
    The grey curves represent the raw data obtained during learning.
    The red curves are the smoothed evolution obtained using Eq.~\ref{eq: smoothed} on the raw datas and with weight $\omega=0.9$.}\label{fig:RL_plots}
\end{figure*}

\section{Appendix C. Results}

We here present the results of our automated approach of photonic-based DIQKD experiments design.
Quantum optical circuits maximizing our designed rewards are presented, as well as their corresponding parameters for different values of efficiency.

\subsection{Discovered quantum optical experiments for DIQKD}

When using the reward given in Eq.\ref{eq: reward 1} to  maximize the key rate in a noise-less scenario, the PPO algorithm converges to the circuit depicted in Fig.~\ref{fig:setup_eff_1}. 
Parameters optimising the key rate for different efficiencies are given in Tab.~\ref{tab:eff_1}.
Similar circuits with extra gates that can freely be discarded ( e.g. a phase shifter on the third mode just before the heralding operation) were also found by the agent.
The occurrence of these similar circuits is simply explained by the reward not punishing for greater circuit depths. Furthermore, adding a "non-contributing" gate can nevertheless increase the reward due to a better parameter optimisation.

In the case of the reward given in Eq.~\ref{eq: reward min} targeting circuits with a high loss tolerance, the PPO algorithm settles on the circuit drawn in Fig.~\ref{fig:best_setup}.
Tab.~\ref{tab:robust} lists the best parameters choice for different values of efficiency.

\sisetup{exponent-mode=scientific, output-exponent-marker = e, round-mode=places,round-precision=4, table-format=1.4}
\begin{table*}[htb]
	\small
    \centering
	\subfloat[][]{
    \begin{tabular}{@{}c|c|cccccccc@{}}
        \toprule
		$\eta$ & key rate & TMS$_{12}^r$ & TMS$_{12}^\theta$ & TMS$_{23}^r$ & TMS$_{23}^\theta$ & TMS$_{13}^r$ & TMS$_{13}^\theta$ & SMS$_2^r$ & SMS$_2^\theta$ \\
        \midrule
        0.0 & \num{0.9137156440435049} & \num{0.049722817215830654} & \num{1.5727154277819255} & \num{1.8764073632879066e-5} & \num{0.9894254331154752} & \num{1.856701335846047e-5} & \num{2.5610126545360554} & \num{-0.05117286572547441} & \num{0.011440006146520468}\\
        0.01 & \num{0.767686824245501} & \num{0.053057525168213125} & \num{1.572791980247873} & \num{0.00037044666908959994} & \num{0.9486749686882656} & \num{0.00036730617439308236} & \num{2.5192272913034697} & \num{-0.05210280697213371} & \num{0.004662270379311472}\\
        0.02 & \num{0.6527097562303206} & \num{0.05817602327926538} & \num{1.5700783994802932} & \num{0.00039093127753258173} & \num{0.9791425378771005} & \num{0.0003886551875928084} & \num{2.549291858704524} & \num{-0.052997503157831756} & \num{-0.00030360716706422173}\\
        0.03 & \num{0.5526583657839196} & \num{0.06372273262796384} & \num{1.5689454916528536} & \num{0.00042080877333334474} & \num{1.0055793496512315} & \num{0.0004187387705064189} & \num{2.5762350970436514} & \num{-0.053323098484885326} & \num{-0.007433092382639303}\\
        0.04 & \num{0.4631271062555493} & \num{0.07050382517622988} & \num{1.5730501478911978} & \num{0.000391120482662727} & \num{0.9869991590599775} & \num{0.00038950892008739034} & \num{2.5583004167843075} & \num{-0.054637094612921655} & \num{0.0073763314859287285}\\
        0.05 & \num{0.3818454061357336} & \num{0.07761142264418251} & \num{1.5706474404665152} & \num{0.0004228669947954165} & \num{1.0026923458886978} & \num{0.0004215177954348078} & \num{2.573535465207127} & \num{-0.05340299350861853} & \num{-0.00038425023527579227}\\
        0.06 & \num{0.3074375199825068} & \num{0.08550331788867509} & \num{1.570870860365692} & \num{0.00044169593857930876} & \num{1.0528150171944795} & \num{0.00044065762826705747} & \num{2.6235941111996004} & \num{-0.051199601182861466} & \num{0.0007461187634882162}\\
        0.07 & \num{0.23904644797744368} & \num{0.09418710759989403} & \num{1.5709843953975242} & \num{0.0005390418793262244} & \num{1.0958348373763576} & \num{0.0005381998231293484} & \num{2.6667227357832255} & \num{-0.04766083556698939} & \num{0.00017860510431416362}\\
        0.08 & \num{0.1762544562691568} & \num{0.10426362591863979} & \num{1.5707691780212856} & \num{0.0007960062434655799} & \num{1.109011082124043} & \num{0.0007951349759846516} & \num{2.679812520814698} & \num{-0.0436297306809708} & \num{-0.00023381535029516875}\\
        0.09 & \num{0.11926196626481445} & \num{0.11654109305761906} & \num{1.5709200564467434} & \num{0.0005839312848007497} & \num{1.1251317911601173} & \num{0.0005829915408562303} & \num{2.6959615654023006} & \num{-0.040095564032252444} & \num{0.0002930597945635898}\\
        0.1 & \num{0.06942244582532187} & \num{0.13162497243649987} & \num{1.5707821576658207} & \num{0.000623103962335954} & \num{1.144162309520832} & \num{0.0006211636182975314} & \num{2.7149738373740187} & \num{-0.03651855876528926} & \num{-0.00019758177371353864}\\
        0.102 & \num{0.060558227278010524} & \num{0.1349450098557854} & \num{1.5707585018512713} & \num{0.00052327903189517} & \num{1.104841830384577} & \num{0.0005214806261334437} & \num{2.6756204497940663} & \num{-0.03559399389404408} & \num{0.00011046840415571578}\\
        0.104 & \num{0.052132005416995586} & \num{0.13831957869586425} & \num{1.570838509049988} & \num{0.0005372365150406038} & \num{1.1237399902587446} & \num{0.0005352115867908647} & \num{2.6945731990640094} & \num{-0.03455160043271094} & \num{-0.00015647773388477416}\\
        0.106 & \num{0.04417876594296988} & \num{0.14173519784970728} & \num{1.570795984902848} & \num{0.0005553791998742446} & \num{1.1331624110867098} & \num{0.0005531220666390443} & \num{2.7039534069068885} & \num{-0.03338063552934783} & \num{1.1993887350053596e-5}\\
        0.108 & \num{0.03673446401658376} & \num{0.1451631198291172} & \num{1.5709719599939036} & \num{0.0004856025982853455} & \num{1.14839525791121} & \num{0.0004835003973559404} & \num{2.719275642911818} & \num{-0.032064383154082596} & \num{0.0003960858230979416}\\
        0.11 & \num{0.029835935177598305} & \num{0.14859466574937116} & \num{1.5708028684071953} & \num{0.0002915696467031814} & \num{1.1589434896667725} & \num{0.00029023848078715765} & \num{2.729830888994221} & \num{-0.03067571562234353} & \num{0.0011961491079098886}\\
        0.112 & \num{0.023520953524101995} & \num{0.15204954880216423} & \num{1.5709255785967369} & \num{0.0003560047563905937} & \num{1.1596446451484608} & \num{0.0003543583893804385} & \num{2.730472614848613} & \num{-0.02890531010016464} & \num{0.002212445255188961}\\
        0.114 & \num{0.017832038305403808} & \num{0.15541713031368432} & \num{1.5708413573159594} & \num{0.0005419738182360385} & \num{1.166925520457237} & \num{0.0005394327598318215} & \num{2.737764830022482} & \num{-0.02732976194882631} & \num{-0.0007121719014161614}\\
        0.116 & \num{0.012816553714857903} & \num{0.15881038048219676} & \num{1.5707930844439608} & \num{0.00026547023905303726} & \num{1.1814948639058773} & \num{0.0002642338658704905} & \num{2.7522358206565096} & \num{-0.02544338052134511} & \num{-0.00047822793100843084}\\
        0.118 & \num{0.008524427376769839} & \num{0.1621073357547439} & \num{1.5707735040394581} & \num{0.0002792909293397914} & \num{1.193727576335724} & \num{0.0002780003512710889} & \num{2.7645041811216307} & \num{-0.023568067954537773} & \num{-0.0007528299341258938}\\
        0.12 & \num{0.005016538228228229} & \num{0.1653552667706421} & \num{1.5710312300450988} & \num{0.0002952309040074982} & \num{1.205906773982519} & \num{0.00029391001647109796} & \num{2.7767573647047468} & \num{-0.02151504338893826} & \num{0.0006600415839161637}\\
        0.121 & \num{0.0035785376988395345} & \num{0.1669390939041563} & \num{1.570917713147816} & \num{0.0003089020471969722} & \num{1.2245585652360282} & \num{0.00030755901392759665} & \num{2.7954372625629733} & \num{-0.020587612570313905} & \num{-0.0008929105485063784}\\
        0.122 & \num{0.002364647591942992} & \num{0.16856203432956793} & \num{1.5710557106023526} & \num{0.0003234384706880014} & \num{1.2424696805967153} & \num{0.0003220596120985984} & \num{2.8132716021150106} & \num{-0.01954058129915898} & \num{0.00266699926583212}\\
        0.123 & \num{0.0013863516710328483} & \num{0.17007275839711888} & \num{1.5711260896316588} & \num{0.0003364655970811522} & \num{1.2582566789855276} & \num{0.0003350608815279443} & \num{2.8292645509200556} & \num{-0.018604427929916314} & \num{0.002119249925479024}\\
        0.124 & \num{0.000656548698126258} & \num{0.17160894911672453} & \num{1.5709909906244712} & \num{0.00035080047364590826} & \num{1.26929557243678} & \num{0.00034936504540068915} & \num{2.8401368373304936} & \num{-0.01798873566590311} & \num{0.007913210983791354}\\
        0.125 & \num{0.00018978825246385167} & \num{0.17307782327531593} & \num{1.5710851695160635} & \num{0.00036625756253021955} & \num{1.2814613728371942} & \num{0.000364747585425352} & \num{2.8522537477426546} & \num{-0.017036459301281} & \num{0.014553110658104187}\\
        0.126 & \num{2.858474128397681e-6} & \num{0.17436523301375673} & \num{1.5708141296925462} & \num{0.00010760292061526622} & \num{1.2278158736822509} & \num{0.00010716453501254617} & \num{2.7979361210256464} & \num{-0.015723791560569942} & \num{0.07660404899100497}\\
        0.1261 & \num{2.4878108340065097e-7} & \num{0.17426100159671037} & \num{1.5730711123190506} & \num{0.00010176682616025924} & \num{1.2163566272273465} & \num{0.00010132967871483578} & \num{2.78698109216216} & \num{-0.01742230883062528} & \num{0.10453563477040777}\\
        \bottomrule
    \end{tabular}
	}
	
    \subfloat[][]{
    \begin{tabular}{@{}c|c|ccccccccc@{}}
        \toprule
        $\eta$ & key rate & $A_0$: PS & $A_0$: D & $A_1$: SMS & $A_1$: D & $B_0$: D & $B_1$: SMS & $B_1$: D & $B_2$: D & $p$ \\
        \midrule
        0.0 & \num{0.9137156440435049} & \num{0.0015706998123422486} & \num{0.2223240376531575} & \num{-0.1949295183449991} & \num{-0.6304693917757955} & \num{-0.16790280938354915} & \num{0.2762770654504241} & \num{0.6847845192733076} & \num{0.2219942102550998} & \num{1.0e-9} \\
        0.01 & \num{0.767686824245501} & \num{0.00023369193425656123} & \num{0.23974632623830228} & \num{-0.19173218870510222} & \num{-0.6283938571004339} & \num{-0.16242911963504506} & \num{0.2852040689312475} & \num{0.7022821053638059} & \num{0.24114785898272503} & \num{-9.28730070638321e-10} \\
        0.02 & \num{0.6527097562303206} & \num{-0.0008048934306972955} & \num{0.26219710725766154} & \num{-0.184281078585384} & \num{-0.622986788359119} & \num{-0.1533367953630917} & \num{0.296632120008165} & \num{0.7266898999305361} & \num{0.264349381566493} & \num{2.323354252523758e-9} \\
        0.03 & \num{0.5526583657839196} & \num{-0.0002354071286790556} & \num{0.2840263257088412} & \num{-0.1777998485651183} & \num{-0.6197614157966962} & \num{-0.1453724702726128} & \num{0.30613514698186206} & \num{0.7489046962490405} & \num{0.28736397404912545} & \num{1.1074422575580379e-8} \\
        0.04 & \num{0.4631271062555493} & \num{0.0008933282207401874} & \num{0.30692116437119443} & \num{-0.17053955446657834} & \num{-0.6175027007275433} & \num{-0.13719574217883967} & \num{0.31607453068736685} & \num{0.771648159424396} & \num{0.3112682196861349} & \num{2.0419157871744076e-7} \\
        0.05 & \num{0.3818454061357336} & \num{5.4204571426773796e-5} & \num{0.33158894781454473} & \num{-0.1626615449317551} & \num{-0.6152166114071772} & \num{-0.12793660691589417} & \num{0.3235869837069408} & \num{0.7947412171146364} & \num{0.33706924968212265} & \num{-6.057099222833355e-6} \\
        0.06 & \num{0.3074375199825068} & \num{6.920407984820129e-5} & \num{0.3572243887988451} & \num{-0.15410041000187233} & \num{-0.6137821294211249} & \num{-0.11854086638856262} & \num{0.3298229818160921} & \num{0.8181252306656703} & \num{0.3639028724801244} & \num{5.8981799070012005e-5} \\
        0.07 & \num{0.23904644797744368} & \num{0.00012173185316537349} & \num{0.38416170653331505} & \num{-0.14479256088679748} & \num{-0.6130811539807546} & \num{-0.10864601095694706} & \num{0.3343341132219597} & \num{0.8416689135852099} & \num{0.3922882776767931} & \num{-0.0003219168573253689} \\
        0.08 & \num{0.1762544562691568} & \num{-3.480308690937066e-5} & \num{0.4130542670675739} & \num{-0.13407338552921041} & \num{-0.6131309143747168} & \num{-0.09785442987137093} & \num{0.33794253089810655} & \num{0.8657890708535518} & \num{0.4230111461720822} & \num{-0.0013199645714249693} \\
        0.09 & \num{0.11926196626481445} & \num{3.912208867092202e-5} & \num{0.4449701571021186} & \num{-0.12094237137025769} & \num{-0.613733280206438} & \num{-0.08550457705605596} & \num{0.341636281072171} & \num{0.8914914420099667} & \num{0.45765863638823} & \num{-0.004852259506123988} \\
        0.1 & \num{0.06942244582532187} & \num{1.9935040990821876e-5} & \num{0.48152381663121135} & \num{-0.1042606867485539} & \num{-0.6144454512422117} & \num{-0.07029022347984129} & \num{0.34499210074197967} & \num{0.9195666798545383} & \num{0.4982602683758703} & \num{0.016935771538582257} \\
        0.102 & \num{0.060558227278010524} & \num{-9.627119167076703e-6} & \num{0.4893963555027374} & \num{-0.10046014668520369} & \num{-0.6145666079484461} & \num{-0.06684005411655773} & \num{0.34542281340576547} & \num{0.9254210007311069} & \num{0.507106111209019} & \num{0.021502754811021364} \\
        0.104 & \num{0.052132005416995586} & \num{5.2223619385928395e-5} & \num{0.4973583755952538} & \num{-0.09674847968481246} & \num{-0.6148567933820477} & \num{-0.06343370992722734} & \num{0.34572874064281106} & \num{0.9312496523708753} & \num{0.516094224794034} & \num{0.027154801280397815} \\
        0.106 & \num{0.04417876594296988} & \num{-2.6836974629667725e-5} & \num{0.5053873446130522} & \num{-0.092925280103019} & \num{-0.6151903346541182} & \num{-0.05991909551641155} & \num{0.34584645944157283} & \num{0.9370551628937265} & \num{0.5251694527490706} & \num{0.034114036468573854} \\
        0.108 & \num{0.03673446401658376} & \num{0.0001164435727344349} & \num{0.5134106412909385} & \num{-0.08927924681595166} & \num{-0.6157264351331588} & \num{-0.05655069602504228} & \num{0.3457178481033893} & \num{0.94271953657722} & \num{0.5342880961963881} & \num{0.04264633449512472} \\
        0.11 & \num{0.029835935177598305} & \num{0.00010766983695990267} & \num{0.5214566796458299} & \num{-0.08556123000475684} & \num{-0.6162735962745164} & \num{-0.05303614131812392} & \num{0.34557604254014335} & \num{0.9483391122470302} & \num{0.5434112900004443} & \num{0.05312376227810184} \\
        0.112 & \num{0.023520953524101995} & \num{-1.937745020801811e-6} & \num{0.5296304897210387} & \num{-0.0817499247732653} & \num{-0.6168312435213615} & \num{-0.04932236247695475} & \num{0.3449790081286149} & \num{0.9539427169120525} & \num{0.5527048502239588} & \num{0.06601325421508838} \\
        0.114 & \num{0.017832038305403808} & \num{5.9934335771380954e-5} & \num{0.5373462272869859} & \num{-0.07839274038562755} & \num{-0.6178736832035487} & \num{-0.046363185881606525} & \num{0.3444754186946367} & \num{0.9591908311498841} & \num{0.5615482412832251} & \num{0.08194652580629365} \\
        0.116 & \num{0.012816553714857903} & \num{-3.7850648876448177e-5} & \num{0.5452379122164293} & \num{-0.07485701532625791} & \num{-0.6188064654847756} & \num{-0.04304540520521788} & \num{0.34363504741302736} & \num{0.9644590407252646} & \num{0.5705832930548705} & \num{0.10191322609109721} \\
        0.118 & \num{0.008524427376769839} & \num{8.770527380338484e-6} & \num{0.5528224408770572} & \num{-0.07177534866792522} & \num{-0.6200628882873218} & \num{-0.04004216133057652} & \num{0.34271173378312736} & \num{0.9693795629111056} & \num{0.5792932792674979} & \num{0.1274109414757529} \\
        0.12 & \num{0.005016538228228229} & \num{0.00010671299231250295} & \num{0.5602677493856949} & \num{-0.06886521934043222} & \num{-0.621530709182866} & \num{-0.03722331863843543} & \num{0.34148268618058336} & \num{0.9740901899395871} & \num{0.5879379565810219} & \num{0.16088067944562626} \\
        0.121 & \num{0.0035785376988395345} & \num{0.00010696173645472561} & \num{0.5638683155415896} & \num{-0.06748310281195473} & \num{-0.6223518218039} & \num{-0.03589504935727844} & \num{0.3409756048920908} & \num{0.9763971725156196} & \num{0.5920436719389436} & \num{0.18197557751133733} \\
        0.122 & \num{0.002364647591942992} & \num{5.671962016932856e-5} & \num{0.5675133524657107} & \num{-0.06612585928200108} & \num{-0.6232575684386877} & \num{-0.03455737790199916} & \num{0.34023333048347426} & \num{0.9786344928813155} & \num{0.5963069566161034} & \num{0.20721000506420378} \\
        0.123 & \num{0.0013863516710328483} & \num{0.0002719267055640455} & \num{0.5709550845814816} & \num{-0.06486732242844513} & \num{-0.6240752625203966} & \num{-0.03327356989923889} & \num{0.3396713155060758} & \num{0.9807526835601155} & \num{0.6002866568014645} & \num{0.23803659656212695} \\
        0.124 & \num{0.000656548698126258} & \num{9.909893146642353e-5} & \num{0.5742383691821337} & \num{-0.06397341321741076} & \num{-0.6252108967903658} & \num{-0.03233365804495943} & \num{0.33941678354591404} & \num{0.9828867764239125} & \num{0.6040667968756441} & \num{0.27752597354351677} \\
        0.125 & \num{0.00018978825246385167} & \num{-0.00019150693639799512} & \num{0.5776180103478544} & \num{-0.06385418751451902} & \num{-0.6266901035770444} & \num{-0.03161109624997573} & \num{0.33896605933127477} & \num{0.9849118705817328} & \num{0.6082119723385566} & \num{0.33330320677004005} \\
        0.126 & \num{2.858474128397681e-6} & \num{-0.0007799723075566727} & \num{0.5803588521837642} & \num{-0.06280274853011852} & \num{-0.627742654057401} & \num{-0.03050058625638148} & \num{0.3375102392225577} & \num{0.9863339626406665} & \num{0.611701087468242} & \num{0.44007795192682914} \\
        0.1261 & \num{2.4878108340065097e-7} & \num{0.00048656634548517656} & \num{0.5797017111214956} & \num{-0.06267569878282321} & \num{-0.6269875470730262} & \num{-0.029760494051951525} & \num{0.33944750702384874} & \num{0.9866304777049558} & \num{0.6110339633185572} & \num{0.47139568286852507} \\
        \bottomrule
    \end{tabular}
    }
    \centering
    \caption{Key rate and optical devices parameters for the circuit depicted in Fig.~\ref{fig:setup_eff_1} for different losses value $\eta$ considered to be the same for all NPNR-photon detectors. Parameters are for (a) state generation parameters (b) measurement settings. Single- and two-mode squeezers have two parameters $r,\theta$ which are given in separate column with name containing $r$ and $\theta$ as exponent. $p$ denotes the noisy pre-processing probability.}
	\label{tab:eff_1}
\end{table*}
\begin{table*}[htb]
    \centering
    \begin{tabular}{@{}c|c|ccccccccc@{}}
        \toprule
        $\eta$ & key rate & TMS$^r$ & TMS$^\theta$ & $A_0$: D$^x$ & $A_1$: SMS & $A_1$: D$^x$ & $B_0$: D$^p$ & $B_1$: D$^p$ & $B_2$: D$^p$ & $p$ \\
        \midrule
       0.0 & \num{0.46009387210564645} & \num{0.7417000768957924} & \num{1.570839830641258} & \num{0.02377876999573749} & \num{-0.3122462926177551} & \num{-0.7999602054055088} & \num{0.2925453939657808} & \num{-0.38842191033081436} & \num{-0.02066158508280264} & \num{3.6645065066271894e-9} \\
        0.01 & \num{0.3924638582073967} & \num{0.7267044973935837} & \num{1.5707901218430695} & \num{0.03350217887156588} & \num{-0.3116830187362099} & \num{-0.7914086656428623} & \num{0.28420636873321775} & \num{-0.39874190215082045} & \num{-0.029351192478674088} & \num{2.102086312261974e-5} \\
        0.02 & \num{0.3355692183226996} & \num{0.7111455591510221} & \num{1.5707912632985428} & \num{0.034711205066284424} & \num{-0.3133535925377775} & \num{-0.7939831125687918} & \num{0.2830388130753859} & \num{-0.4008288643481947} & \num{-0.02997778814417227} & \num{0.00016367314057320873} \\
        0.03 & \num{0.284114881611241} & \num{0.6931088853766729} & \num{1.570830504764672} & \num{0.03480560825067238} & \num{-0.31546054416536146} & \num{-0.7975897135149203} & \num{0.28216311552880213} & \num{-0.4015189335485504} & \num{-0.029529167168821004} & \num{-0.0005862341466553439} \\
        0.04 & \num{0.23708460097593353} & \num{0.6719509192475683} & \num{1.5708231127778154} & \num{0.03432481704977402} & \num{-0.31768420374513223} & \num{-0.8011785024075032} & \num{0.28087659200178905} & \num{-0.40114780424521246} & \num{-0.028495968877045048} & \num{0.0014921127552317492} \\
        0.05 & \num{0.1941927581225208} & \num{0.6469347749832935} & \num{1.5707699852855503} & \num{0.03343332095563525} & \num{-0.32004919684270666} & \num{-0.8043443662829997} & \num{0.27863398718588595} & \num{-0.3997538983056988} & \num{-0.02699724473316053} & \num{-0.0031228772714421036} \\
        0.06 & \num{0.1554552938093603} & \num{0.6177000482591506} & \num{1.5708080402964546} & \num{0.032268918938234796} & \num{-0.32235932625545993} & \num{-0.806444467969606} & \num{0.2749489872947484} & \num{-0.3972722080058082} & \num{-0.02518567967328743} & \num{0.00575116750904362} \\
        0.07 & \num{0.12103088228206094} & \num{0.5838392720398357} & \num{1.570779012306884} & \num{0.030851100511420294} & \num{-0.32452265626107435} & \num{-0.8072027880116769} & \num{0.26936271685463475} & \num{-0.3932832528669663} & \num{-0.0230426654739801} & \num{0.00963533699986678} \\
        0.08 & \num{0.09112026719657385} & \num{0.5453297125694256} & \num{1.5707544173984824} & \num{0.029309057595758473} & \num{-0.32625741818669574} & \num{-0.8057921494979312} & \num{0.2611303048673255} & \num{-0.3876684320481515} & \num{-0.020728547774400207} & \num{0.015053971242533937} \\
        0.09 & \num{0.06587876808081983} & \num{0.5024424985687558} & \num{1.5708315737780572} & \num{0.02757323746676478} & \num{-0.3271411899371896} & \num{-0.8017645464701091} & \num{0.24983548010447298} & \num{-0.3799018580857866} & \num{-0.018173851984017098} & \num{0.022276778338882093} \\
        0.1 & \num{0.04533808875524736} & \num{0.45579488776630017} & \num{1.5708053956692622} & \num{0.025683760813348517} & \num{-0.32673963462730987} & \num{-0.7943736114343984} & \num{0.2349135802709727} & \num{-0.36943765156832536} & \num{-0.015524834845498442} & \num{0.03154064435826958} \\
        0.11 & \num{0.029349988270456873} & \num{0.4061030925435241} & \num{1.570820839290217} & \num{0.023558623989765503} & \num{-0.32433245078371936} & \num{-0.7828962171783246} & \num{0.21601468004039662} & \num{-0.3556766443544208} & \num{-0.012804241760518661} & \num{0.043135729087237014} \\
        0.12 & \num{0.017566059072469442} & \num{0.3541561743213891} & \num{1.5708629093913724} & \num{0.021184324831322522} & \num{-0.31911897046329407} & \num{-0.7661916095080588} & \num{0.19294936203642704} & \num{-0.33788851618722304} & \num{-0.010095120513637994} & \num{0.0575837498064742} \\
        0.13 & \num{0.00945780491499737} & \num{0.3006994382305477} & \num{1.5706978044473856} & \num{0.01841540685245236} & \num{-0.31025755970588703} & \num{-0.74333119313314} & \num{0.1659026894290832} & \num{-0.31523378740212715} & \num{-0.007461662111265951} & \num{0.07573615879348464} \\
        0.14 & \num{0.004367574726555135} & \num{0.24608771828782502} & \num{1.570796614081143} & \num{0.015200055413917115} & \num{-0.29614212055074507} & \num{-0.7121564835017425} & \num{0.13523977020163064} & \num{-0.28646867195150816} & \num{-0.005057933981979447} & \num{0.09917055111819491} \\
        0.15 & \num{0.001573089550898188} & \num{0.1904258184565963} & \num{1.5707237510888146} & \num{0.011506510490320187} & \num{-0.2752444103453177} & \num{-0.6703674279001478} & \num{0.10179864972504446} & \num{-0.2496440709528563} & \num{-0.00297204384519507} & \num{0.1311466977386621} \\
        0.16 & \num{0.0003492137092130365} & \num{0.1329128331243102} & \num{1.570762933583878} & \num{0.007265664249142323} & \num{-0.24497761392360987} & \num{-0.6138824262580264} & \num{0.06676169659115623} & \num{-0.2007869202209524} & \num{-0.0013031980558572492} & \num{0.18007795418383865} \\
        0.161 & \num{0.00028648431374311834} & \num{0.12690738525054207} & \num{1.5707849307359145} & \num{0.006814029914285867} & \num{-0.24117788689298722} & \num{-0.6068253799656692} & \num{0.06311200464763786} & \num{-0.19485335515030855} & \num{-0.001165744270708223} & \num{0.18631097973002614} \\
        0.162 & \num{0.00023197496458049471} & \num{0.12105387834704937} & \num{1.5703988974789527} & \num{0.006390348595795667} & \num{-0.23769087329645464} & \num{-0.6003954639305813} & \num{0.0597131026449628} & \num{-0.18875409456945927} & \num{-0.0010167494399187034} & \num{0.1932980597510728} \\
        0.163 & \num{0.00018505679440439238} & \num{0.11516688530073484} & \num{1.57073124132386} & \num{0.005959400613504912} & \num{-0.23351196430638552} & \num{-0.593029246611218} & \num{0.05621698145116907} & \num{-0.18270274895929164} & \num{-0.0009092938939685616} & \num{0.2008980740358195} \\
        0.164 & \num{0.00014510957790903056} & \num{0.1090388481817058} & \num{1.5707201327896672} & \num{0.005519892497724053} & \num{-0.22939305408135982} & \num{-0.5854641575535133} & \num{0.05268331401418812} & \num{-0.1761412029016295} & \num{-0.0008122758253907174} & \num{0.20849686498343156} \\
        0.165 & \num{0.00011152507841294401} & \num{0.10275828657890512} & \num{1.5710835177905167} & \num{0.005026995640331716} & \num{-0.2250717974546318} & \num{-0.5775924510798233} & \num{0.04903551750079719} & \num{-0.16902587305721248} & \num{-0.0006903102147790759} & \num{0.21760983976054776} \\
        0.166 & \num{8.370598813145502e-5} & \num{0.09628686859600435} & \num{1.5706724288209455} & \num{0.004563068899503225} & \num{-0.2205580917278957} & \num{-0.5696180899660244} & \num{0.04541445180305898} & \num{-0.16135161916440902} & \num{-0.0005931641602373663} & \num{0.2258650447644764} \\
        0.167 & \num{6.106414754258882e-5} & \num{0.09001455791420655} & \num{1.5710210919898335} & \num{0.004125573605417319} & \num{-0.2159530541640542} & \num{-0.5606839521163839} & \num{0.04183788086302287} & \num{-0.1540255158324142} & \num{-0.00050247707556567} & \num{0.23633017477963913} \\
        0.168 & \num{4.301978463838729e-5} & \num{0.0838747339798898} & \num{1.5707816966436443} & \num{0.00371925868620828} & \num{-0.211101621170394} & \num{-0.5522330553895619} & \num{0.03843930846271649} & \num{-0.14619637021041487} & \num{-0.0004485479016983688} & \num{0.2474669092689109} \\
        0.169 & \num{2.9019338242508574e-5} & \num{0.07683741058534747} & \num{1.5707780847291373} & \num{0.003194394554459478} & \num{-0.20594058161905843} & \num{-0.5419914664596466} & \num{0.03459992572672523} & \num{-0.1372152656419072} & \num{-0.0003327202943894967} & \num{0.25949879831102934} \\
        0.17 & \num{1.8506913159410665e-5} & \num{0.07004745525814918} & \num{1.5698651906824064} & \num{0.002811291293992149} & \num{-0.1998289445239373} & \num{-0.5316036654409902} & \num{0.030949804539799036} & \num{-0.1280819289893427} & \num{-0.0003009599953402368} & \num{0.2718584040941959} \\
        0.171 & \num{1.0951958658189653e-5} & \num{0.06274961000782792} & \num{1.5714929368730923} & \num{0.0023506346850121883} & \num{-0.19392755000361842} & \num{-0.5211856922683729} & \num{0.027192570902214743} & \num{-0.11761055495462057} & \num{-0.00016297767514758594} & \num{0.28703067022699186} \\
        0.172 & \num{5.832911937564411e-6} & \num{0.05578086837262367} & \num{1.5714742227759038} & \num{0.0018728407625518775} & \num{-0.18825064518325046} & \num{-0.510902320660556} & \num{0.02372172562006009} & \num{-0.10696657210224317} & \num{-0.0001289177386867914} & \num{0.306088723238542} \\
        0.173 & \num{2.647582093451284e-6} & \num{0.04793240602619408} & \num{1.5706289263391604} & \num{0.0015119668549804476} & \num{-0.1783713333092117} & \num{-0.4946898242574739} & \num{0.019848683420953295} & \num{-0.095389380711792} & \num{-9.721313196711933e-5} & \num{0.32487749901424606} \\
        0.174 & \num{9.19168633206624e-7} & \num{0.03850447860781037} & \num{1.5710967614554512} & \num{0.0010040826128650557} & \num{-0.17120308462642608} & \num{-0.4808685847846702} & \num{0.01546016737438238} & \num{-0.07919940823162692} & \num{-6.423261794932558e-5} & \num{0.35135109310686274} \\
        0.175 & \num{1.6980974582025965e-7} & \num{0.03205192147316328} & \num{1.572209243076593} & \num{0.0008032324465915477} & \num{-0.1703794848068441} & \num{-0.473689340434537} & \num{0.012677544214357815} & \num{-0.06697068246784109} & \num{1.102520335303405e-5} & \num{0.3950502382836251} \\
        0.1756 & \num{1.0870893296655026e-8} & \num{0.031382360419748935} & \num{1.5694489126080222} & \num{0.0006620202042259442} & \num{-0.16206032439412887} & \num{-0.45882170376078923} & \num{0.01208813771667306} & \num{-0.06812949239621328} & \num{-0.00019527681105952383} & \num{0.43388371177111495} \\
        \bottomrule
    \end{tabular}
	\caption{Key rate and optical devices parameters for the circuit depicted in Fig.~\ref{fig:best_setup} for different losses value $\eta$ considered to be the same for all NPNR-photon detectors.}
	\label{tab:robust}
	
\end{table*}

\end{widetext}
\end{document}